\documentclass[aps,pra,twocolumn,showpacs,superscriptaddress,10p,longbibliography]{revtex4}
\usepackage{graphicx}
\usepackage{epsfig}
\usepackage{dcolumn}
\usepackage{amsmath}

\usepackage[sort&compress]{natbib}

\usepackage{bm}
\usepackage{bbm}
\usepackage{ulem}
\usepackage[colorlinks, citecolor=red]{hyperref}
\usepackage{mathtools}
\usepackage{color}
\usepackage{soul}
\usepackage{centernot}
\usepackage[up]{subfigure}
\usepackage{dsfont}	
\usepackage{relsize}
\usepackage{epstopdf}
\usepackage{mathrsfs}
\usepackage{braket}
\begin{document}
\title{A Nonlinear Journey from Structural Phase Transitions to Quantum Annealing}
\author{Mithun Thudiyangal\href{https://orcid.org/0000-0003-4341-6439}{\includegraphics[scale=0.05]{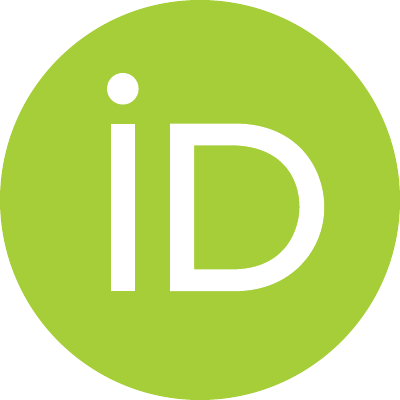}}}
\affiliation{Department  of  Atomic and Molecular Physics,  Manipal Academy of Higher Education, Manipal 576 104, India}
\author{Panayotis G. Kevrekidis\href{https://orcid.org/0000-0002-7714-3689}{\includegraphics[scale=0.05]{orcidid.pdf}}}
\affiliation{Department of Mathematics and Statistics, University of Massachusetts, Amherst MA 01003-4515, USA} 
\author{Avadh Saxena\href{https://orcid.org/0000-0002-3374-3236}{\includegraphics[scale=0.05]{orcidid.pdf}}}
\author{Alan R. Bishop}
\affiliation{Center for Nonlinear Studies and Theoretical Division, Los Alamos National Laboratory, Los Alamos, New Mexico 87545, USA}
 \begin{abstract}
 Motivated by an exact mapping between equilibrium properties of a 1-dimensional chain of quantum Ising spins in a transverse field (the transverse field Ising (TFI)  model) and a 2-dimensional  classical array of particles in double-well potentials (the ``$\phi^4$ model'') with weak inter-chain coupling, we explore connections 
 between the driven variants of the two systems. We argue that coupling between the fundamental topological solitary waves in the form of kinks between neighboring chains in the classical $\phi^4$ system is the analogue of the competing effect of the transverse field on spin flips in the quantum TFI model. As an example application, we mimic simplified measurement protocols in a closed quantum model system by studying the classical $\phi^4$ model subjected to periodic perturbations. This reveals memory/loss of memory and coherence/decoherence regimes, whose quantum analogues are essential in annealing phenomena. In particular, we examine regimes where the topological excitations control the thermal equilibration following perturbations. This paves the way for further explorations of the analogy
   between lower-dimensional linear quantum and higher-dimensional classical nonlinear systems.
 \end{abstract}
\maketitle
%

\section{Significance Paragraph}
The transverse field Ising (TFI) model  is the basis for quantum annealing architectures in, e.g., D-Wave machines so that the mapping explored herein holds promise for a deeper understanding of such processes. On the other hand, the $\phi^4$ model is a prototypical
nonlinear classical field theory at the epicenter of the analysis of solitary wave dynamics. It has also been
centrally impacted by the fundamental contributions of David Campbell and his collaborators in connection to 
the interactions of the principal (topological) solitary waves, namely the kink structures.
Our aim herein is to exploit the connection between the (lower-dimensional) 
TFI and the (higher-dimensional) $\phi^4$ model in order to understand fundamental features of the former (including
memory and decoherence effects), by studying the classical and more tractable (albeit higher-dimensional and nonlinear) dynamics of the latter. This study is an initial step toward understanding analogies between information processing in quantum linear and classical nonlinear systems.
\section{Introduction}
A distinctive feature of nonlinear science is the ubiquity of phenomena and classes of nonlinear equations that exhibit them –- ubiquity in the spectrum of disciplines and physical scales. From this perspective and in recognition of this celebratory volume for David K. Campbell (DKC), we discuss here a journey of over 50 years involving the double well $\phi^4$ equation. One of us (ARB) shared an interest with DKC, although with different discipline motivations,  in this model nonlinear equation since first meeting in the Aspen Physics Center, while another (PGK) worked with DKC a few decades later on a variant of the $\phi^4$ model inspired by 
ultracold atomic gases in a book chapter within~\cite{dkc}. These snippets, as well as this chapter, underpin a 
journey from modeling classical structural phase transitions in materials such as perovskites \cite{venkataraman1979soft,howard2005structures} to modern special purpose 
quantum systems and computing architectures for studying quantum annealing, for example D-Wave machine based on Josephson junction (SQUID) technology; see for a recent example,
e.g.~\cite{PhysRevB.107.075412}.

 The $\phi^4$ Hamiltonian was employed in the early 1970s to model certain structural phase transitions because new generations of inelastic neutron scattering experiments were becoming capable of resolving much lower frequency scales~\cite{Dodd}. In particular, they suggested the phenomena of “soft modes” (softening of phonon frequency around a characteristic temperature $T_s$) and ``central peaks'' (scattering centered around zero frequency) \cite{koehler1975molecular,venkataraman1979soft,howard2005structures}. The excitations captured by the $\phi^4$ model gave explanations for both of these phenomena~\cite{PhysRevB.34.6295}. As temperature T increases, atomic oscillations (or unit cell rotations) change from (non)linear vibrations in either of the double wells to (non)linear oscillations centered around the maximum of the double well with large amplitude vibrations visiting both wells. The vibration frequency decreases as $T_s$ is approached from above or below. However for $T\lesssim T_s$, $\phi^4$ excitations in the form of thermal transitions from one well to the other also become thermodynamically relevant. In one dimension (1d) these are soliton-like domain walls, {i.e.,} kinks (K) and antikinks ($\bar{\rm K}$). The diffusive dynamics of the kinks and antikinks (see, e.g.,~\cite{PhysRevLett.84.1070}) produces a central scattering intensity peak, with decreasing frequency width as $T$ decreases and the density of K and $\bar{\rm K}$ decreases. The kink diffusion has been analyzed extensively in the context of nonlinear science, including K$\bar{\rm K}$ nucleation, annihilation~\cite{PhysRevLett.84.1070,1624301} and bound states (transient breathers)~\cite{bishop1989global,rasmussen2000discrete,Flach_1998,Flach_2008}. In 1d there is an Ising symmetry breaking critical temperature $T_c$ only at $T = 0;$ for $T>T_c$ the average displacement is zero.  Previous mean-field (self-consistent phonon) approximations predicted $T_c$ and $T_s$ were the same but nonlinear analysis was able to clarify the difference; indeed, central peaks have appeared subsequently in more exotic many-body scenarios but with similar origins. A closely related potential to the $\phi^4$, the double-Gaussian, explicitly separates the Ising criticality as $T \rightarrow T_c$ from the Gaussian fluctuations for $T<T_s$ \cite{Baker1982critical}.

 Here we return to an example of the above scenario where $T_c$ is finite but well-separated from the higher temperature $T_s$. 
Namely, we consider a 2d system of weakly  coupled $\phi^4$ chains. This situation is directly relevant to structural transitions in certain low-dimensional materials \cite{Baker1982critical}.  However, our interest here is that the equilibrium 2d classical Hamiltonian, when the chain-chain coupling is weak, can be exactly mapped to a quantum 1d model, the transverse field Ising (TFI) model \cite{pfeuty1970one}. This classical d-dimensional nonlinear to quantum, linear
(d-1)-dimensional model mapping is interesting as an early example of supersymmetry and quantum phase transitions, whose studies have become extensive more recently~\cite{bernal2022coherent}. However, for our specific interest here, the TFI has been central during the last decade as the basis for ``quantum computing'' architectures such as D-Wave, designed to accelerate annealing~\cite{king2022}. Proposed quantum computing advantages over classical computing are based on linear
quantum evolution. However, it is important to note that classical, but nonlinear, systems can avoid some of the disadvantages of classical linear computing. Indeed there are classical analogs of many phenomena of quantum linear systems, such as entanglement, (de)coherence, memory, superposition, interaction at a distance, a timely theme that we have also considered in earlier 
work; see, e.g.,~\cite{PhysRevE.105.034210,Lloyd2000quantum} and also references therein. In some respects this is not surprising since most classical nonlinear equations arise from slaving between two or more fields – as in semiclassical Bardeen–Cooper–Schrieffer (BCS) or Bose–Einstein condensation (BEC) 
models. At a deeper level, the d-dimensional (classical, nonlinear) to (d-1)-dimensional (quantum, linear) model connection motivating our study here is of course linked to the enduring issues of classical to quantum transitions, but the latter more general theme is beyond the scope of the present study.

 Among the many phenomena shared by the classical weakly-coupled $\phi^4$ chains and the quantum TFI, we focus here on the $\phi^4$ regime where slow relaxation is prevalent because this is most relevant to annealing protocols. This regime ($T_c \lesssim T_s/3$) is where the topological 
 structure, i.e., K$\bar{\rm K}$ (kink-antikink) dynamics is responsible for the relaxation after the perturbations
 emulating (for our purposes) the process of measurement. In our 2d setting, the coupling between 1d chains impedes the 1d chain kink diffusion by the effects of  kinks (or antikinks) interacting and aligning between chains \cite{gro1992langevin,horovitz1977solitons}. For structural phase transitions this is the mechanism by which 2d cluster boundaries form between clusters of particles residing in one well inside and in the other well outside the cluster.
Kink-kink coupling between chains introduces a new timescale by transiently pinning a kink (analogous to
thermal trapping/detrapping by disorder) and allows time for kink alignment across multiple chains. 
Scaling theory, in the form of real-space (block) Renormalization Group analysis can predict the probability distribution of cluster sizes asymptotically as $T \rightarrow T_c$, for example two delta functions (the two wells) in 1d. However,
the dynamics of the clusters for $T_c< T<T_s$ is intricate – inter-chain kink alignment is a dynamic process of binding and unbinding, clusters can move and change shape by secondary K and $\bar{\rm K}$ nucleating on cluster boundaries and propagating around the boundary, etc. In terms of the mapping to a 1d quantum TFI situation, the inter-chain coupling maps to the transverse field frustrating the pure Ising spin flips (analogous to the classical K, $\bar{\rm K}$). In both cases, the topological K, $\bar{\rm K}$ excitations control the annealing regimes. This role of topological excitations is, in fact, a feature of many nonlinear systems. For example, similar phenomena were earlier identified~\cite{PhysRevE.105.034210} 
 in studies of the far-reaching (in terms of its applications) complex Ginzburg-Landau model~\cite{aranson}. In the latter, both vortices and domain walls are the relevant topological excitations. More generally, topological defect-controlled dynamics and relaxation commonly arise in such diverse systems as frustrated Josephson Junction arrays \cite{gro1992langevin}, dislocation patterns controlling stress-strain relationships \cite{LeSar2014simulations}, and monopoles in spin ice \cite{Nisoli2020equilibrium}.

\section{The Model: Weakly coupled $\phi^4$ chains}\label{model}
      \begin{figure}[!htbp]  
 \includegraphics[scale=1.0,width=0.8\columnwidth]{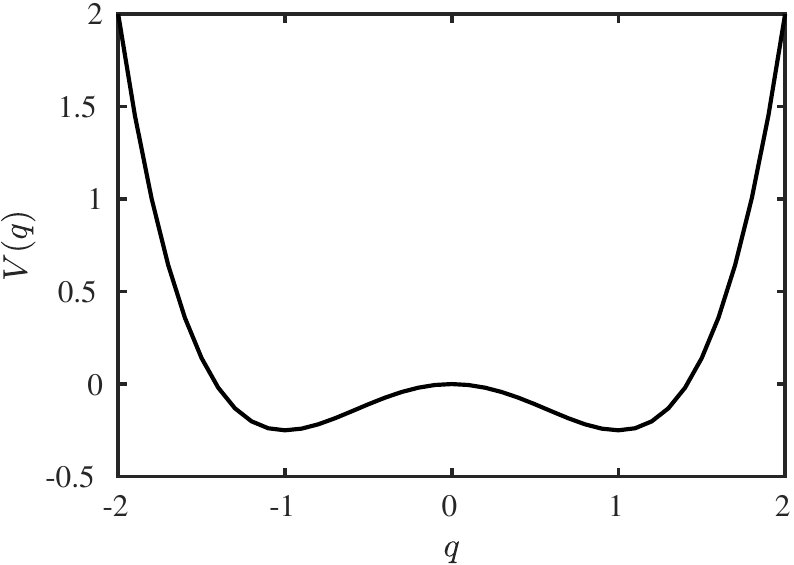} 
 \caption{\label{fig:pot}{\footnotesize {} The non-interacting potential energy function $V(q)$ as a function of the particle displacement $q$ for the fixed parameters $A=-1$, $B=1$.}}
   \end{figure}
We consider the two-dimensional (2d) weakly coupled $\phi^4$ chains given by the Hamiltonian  \cite{PhysRevB.34.6295,PhysRevLett.42.937,PhysRevB.12.2824,bishop1989global,rasmussen2000discrete};
\begin{align}\label{eq:refhamil1}
 \mathcal{H}  = & \sum_{l=1}^L \sum_{m=1}^M \Bigg[\frac{p_{l,m}^2}{2}+V(q_{l,m}) + W_{||}\big(q_{l+1,m}- q_{l,m}\big) \nonumber\\
  & + W_{\perp}\big(q_{l,m+1}- q_{l,m}\big) \Bigg],
\end{align}
where $q_{l,m}$ represents the displacement of a particle at the $l$-th position along the $m$-th chain, and
\begin{align}\label{eq:potential1}
 V(q)  = & A \frac{q^2}{2} +B\frac{q^4 }{4},
\end{align}
denotes the non-interacting potential energy function. Additionally, 
\begin{align}\label{eq:potential2}
  W_{||}(q)= & \frac{C_{||}}{2} q^2~~\text{and}~~
  W_{\perp}(q) = \frac{C_{\perp}}{2}q^2,
\end{align}
respectively. The Hamiltonian Eq.~\eqref{eq:refhamil1} can be rewritten as:
\begin{align}\label{eq:refhamil2}
 \mathcal{H}  = \mathcal{H}_K +\mathcal{H}_V, 
 \end{align}
  where
  \begin{align}\label{eq:Hk}
     \mathcal{H}_K=\sum_{l=1}^L \sum_{m=1}^M \frac{p_{l,m}^2}{2},
  \end{align}
  and
    \begin{align}\label{eq:Hv}
  \mathcal{H}_V &=\sum_{l=1}^L \sum_{m=1}^M \Bigg[\frac{A}{2} q_{l,m}^2  +\frac{B}{4} q_{l,m}^4 \\ \nonumber 
     &+\frac{C_{||}}{2}  \big(q_{l+1,m}- q_{l,m}\big)^2 + \frac{C_{\perp}}{2}\big(q_{l,m+1}- q_{l,m}\big)^2\Bigg].
 \end{align}

In Eq.~\eqref{eq:potential1} $A<0$ and $B>0$  are the parameters of the double well potential, shown in Fig.~\ref{fig:pot}; while, $C_{||}$ and $C_{\perp}$ are the intra-chain and inter-chain coupling constants. We fix $A=-1$, $B=1$ for the numerical simulation of the model Eq.~(\ref{eq:refhamil2}). Moreover, we consider that the interaction within each chain is much stronger than the inter-chain interaction ({i.e.}, $C_{||} \gg C_{\perp}$)

The double well of the potential energy function, Eq.~\eqref{eq:potential1}, shown in Fig. \ref{fig:pot},
has two minima positioned at $\pm1$ with magnitude $-1/4$. We derive the following equations of motion from the Hamiltonian of Eq.~\eqref{eq:refhamil2}: 

\begin{align}
  \ddot{q}_{l,m}
  = &  q_{l,m} -  q_{l,m}^3  \hspace{1cm}\nonumber\\
 & + C_{\perp} \big( q_{l+1, m}-2 q_{l,m}+ q_{l-1,m})
 \nonumber\\
 &  +C_{||}\big(q_{l, m+1}- 2 q_{l,m}+ q_{l, m-1} \Big).
 \label{eq:phi4}
\end{align}

Eq.~\eqref{eq:phi4} possesses one conserved quantity, the total energy $\mathcal{H}$.
Corresponding to the conserved quantity, we define energy per particle (energy density) $h = \frac{\mathcal{H}}{L\times M}$.
 %
             \begin{figure}[!htbp]  
  \includegraphics[scale=1.0,width=0.4\textwidth]{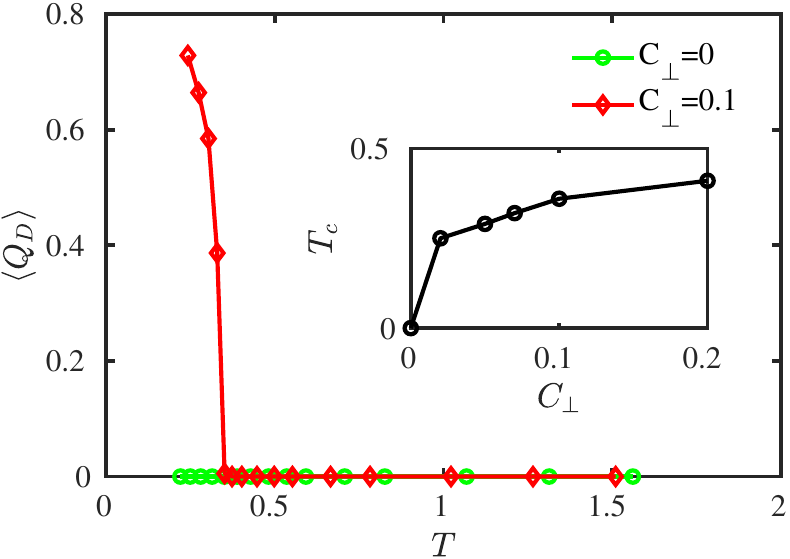} 
 \caption{\label{fig:pav}{\footnotesize {} The average displacement $\langle Q_D \rangle$ for inter-chain coupling strengths $C_{\perp}=0$ (green circles)  and $C_{\perp}=0.1$ (red diamonds) as a function of temperature $T$ for the fixed intra-chain coupling strength $C_{||}=1$. The inset shows the critical temperature $T_c$ associated with the structural phase transition for different $C_{\perp}$.}}
   \end{figure}
   
                      \begin{figure}[!htbp]  
  \includegraphics[scale=1.0,width=0.4\textwidth]{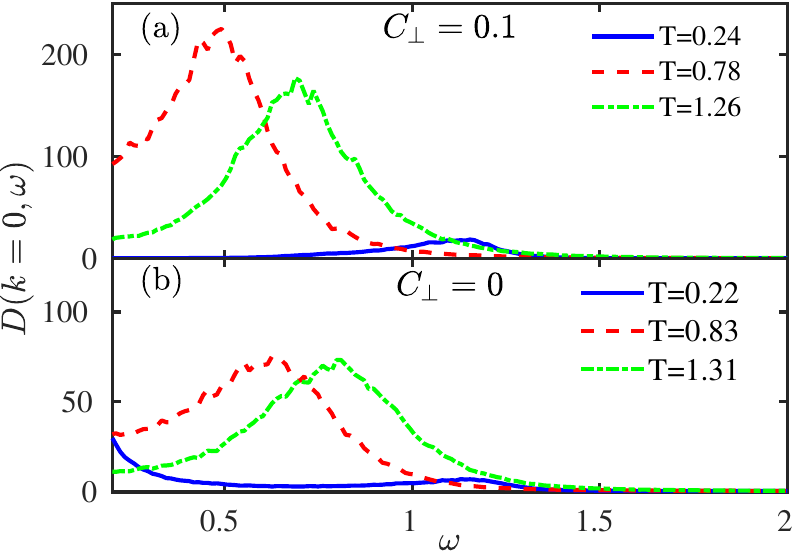}\\
\includegraphics[scale=1.0,width=0.4\textwidth]{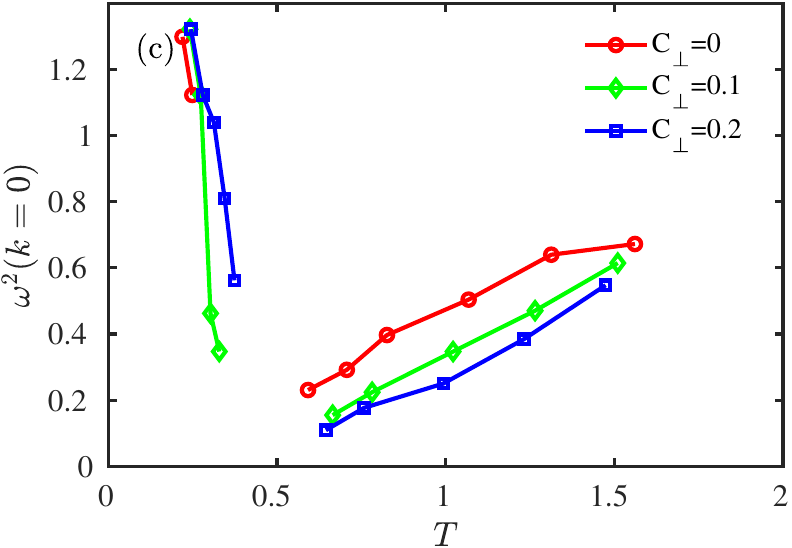} 
 \caption{\label{fig:wsp}{\footnotesize {}  (a-b) The spectral function $D(k=0,\omega)$ for different values of $C_{\perp}$ at different temperatures and (c) square of the soft-mode frequency $\omega_s^2(k=0)$ at different temperatures (c). }}
   \end{figure}

         \begin{figure*}[!htbp]  
  \includegraphics[scale=1.0,width=0.31\textwidth]{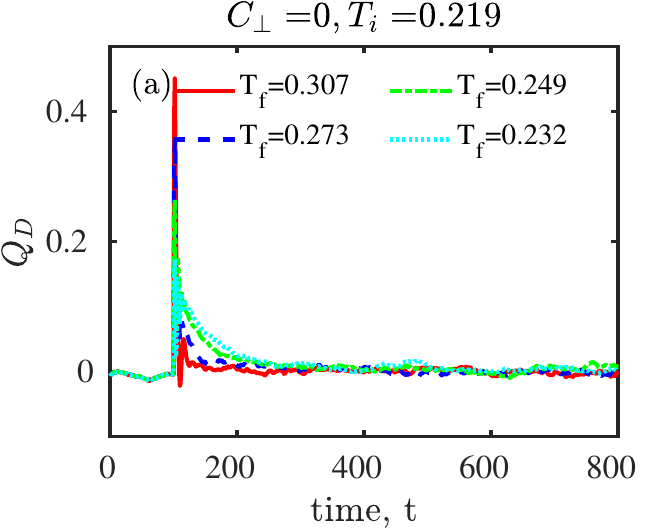}
    \includegraphics[scale=1.0,width=0.31\textwidth]{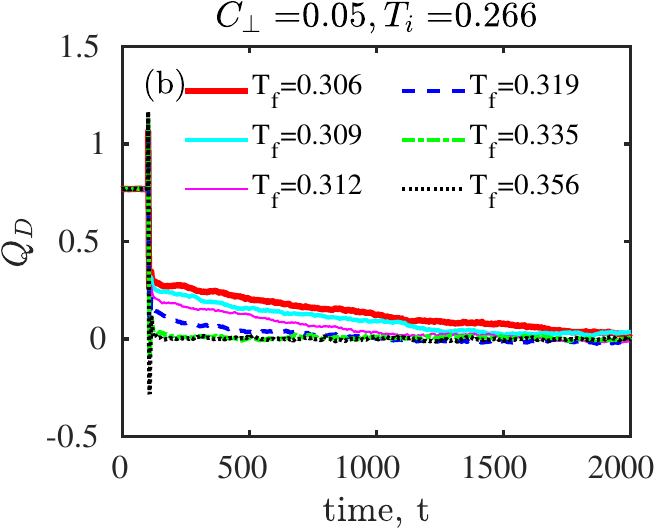}
      \includegraphics[scale=1.0,width=0.31\textwidth]{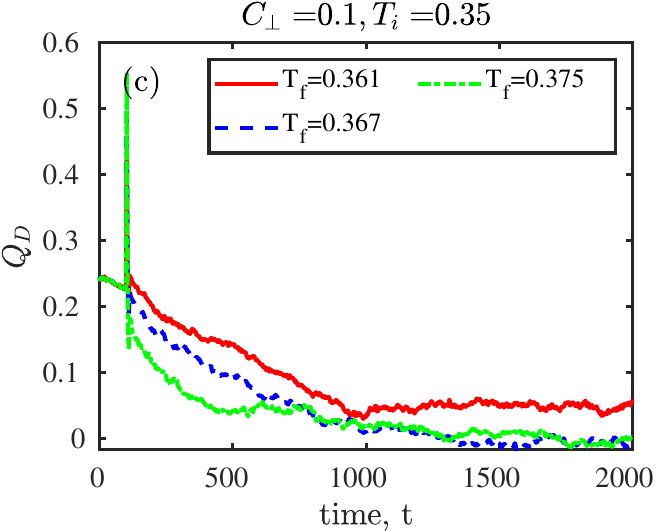}
\caption{{\footnotesize {} Average displacement $Q_D$ for $C_{\perp}=(0,0.05,0.1)$,  (a-c). The other parameters are $L=256$ and $M=10$.  }}
  \label{fig:skickd}
   \end{figure*}

            \begin{figure}[!htbp]   
    \includegraphics[scale=1.0,width=0.4\textwidth]{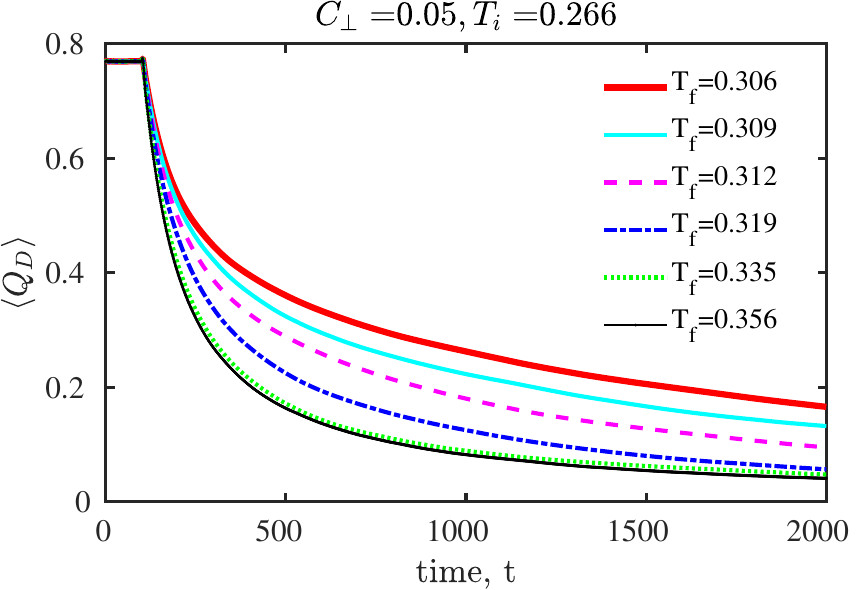}
\caption{{\footnotesize {} Time-averaged displacement $\langle Q_D \rangle$ for $C_{\perp}=0.05$. The other parameters are $L=256$ and $M=10$.  }}
  \label{fig:skickavgd}
   \end{figure}

         \begin{figure}[!htbp]  
  \includegraphics[scale=1.0,width=0.4\textwidth]{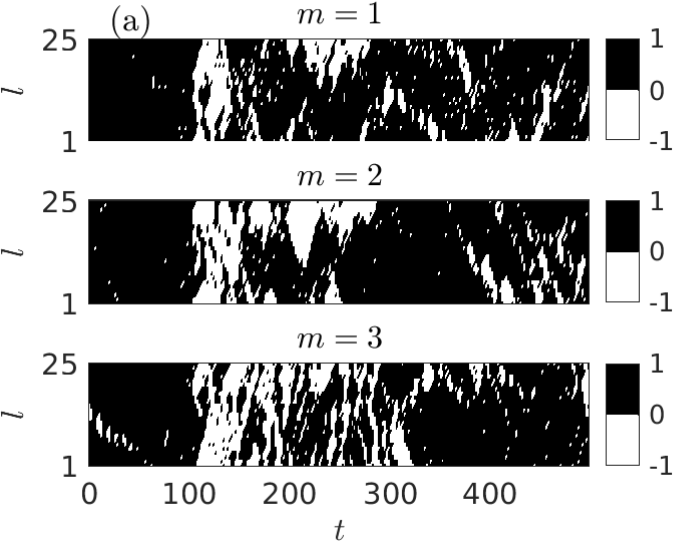}\\
 \includegraphics[scale=1.0,width=0.4\textwidth]{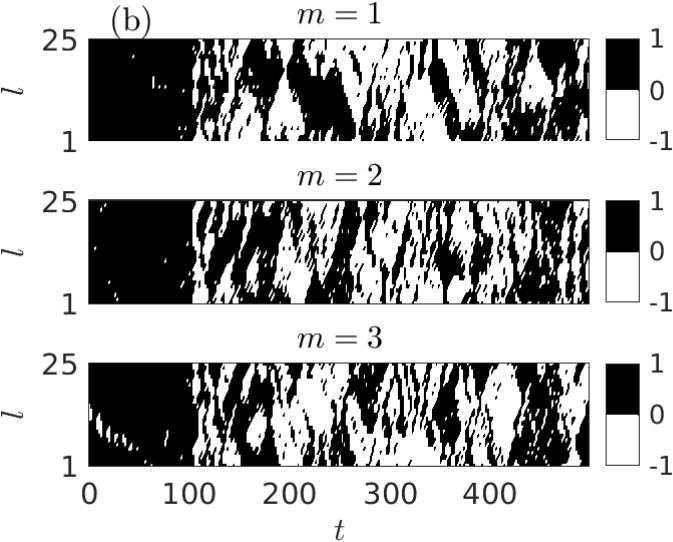}
\caption{{\footnotesize {} Soliton detector: The triplets represent the cases of $C_{\perp}=0.05$ for two different kick strengths (a) $P_0=8$ and (b) $P_0=11$. Each of the triplets contains three chains. The other parameters are $C_{||}=1$, $L=256$ and $M=10$. 
}}
  \label{fig:evolutionsingle}
  \end{figure}

         \begin{figure}[!htbp]  
  \includegraphics[scale=1.0,width=0.4\textwidth]{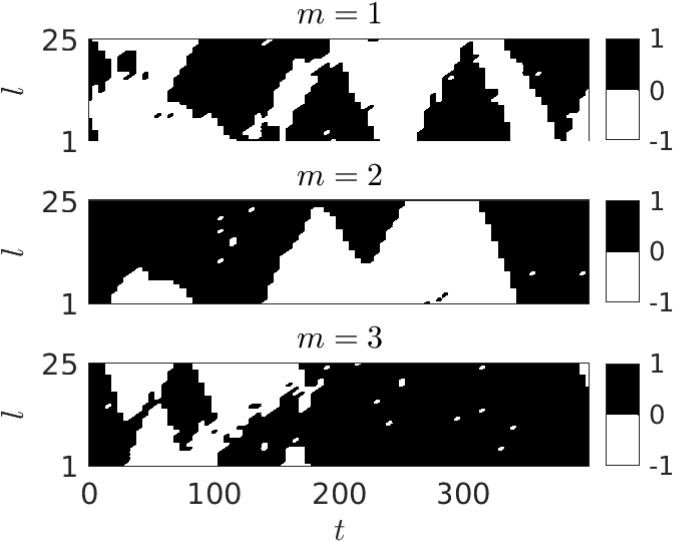}\\
\caption{{\footnotesize {} Soliton detector: The triplets represent the cases of $C_{\perp}=0$ for a kick strength. Each of the triplets contains three chains. The specific parameters are $C_{||}=1$, $L=256$ and $M=10$.  }}
  \label{fig:evolutioncperp0}
   \end{figure}

        \begin{figure}[!htbp]  
    \includegraphics[scale=1.0,width=0.23\textwidth]{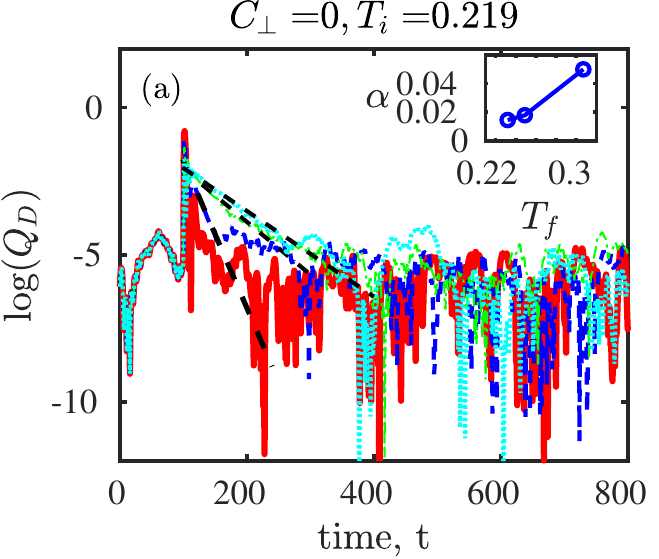}
    \includegraphics[scale=1.0,width=0.23\textwidth]{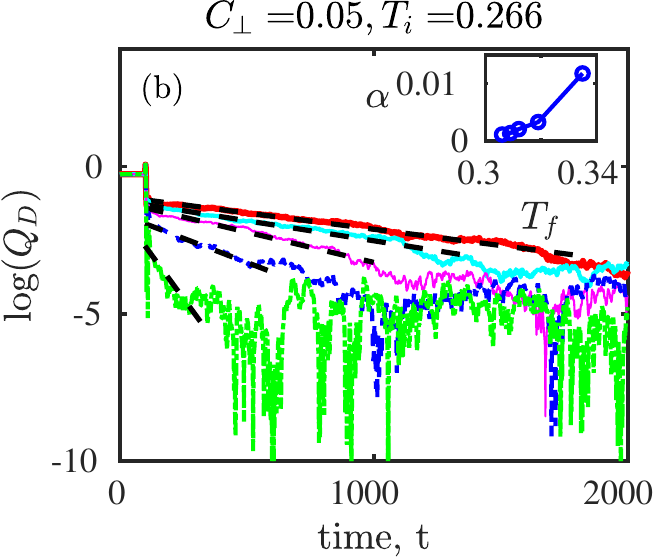}
\caption{{\footnotesize {} Average displacement $Q_D$ for (a) $C_{\perp}=0$ and (b) $C_{\perp}=0.05$. Solid lines represent the fit $\sim e^{-\alpha t}$ and inset represents the values of $\alpha$. The other parameters are $L=256$ and $M=10$.  }}
  \label{fig:skickdfit}
   \end{figure}

           \begin{figure}[!htbp]   
    \includegraphics[scale=1.0,width=0.23\textwidth]{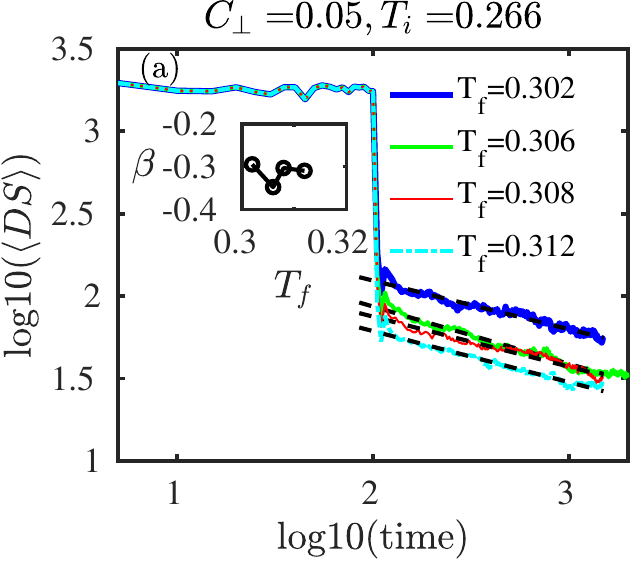}
       \includegraphics[scale=1.0,width=0.23\textwidth]{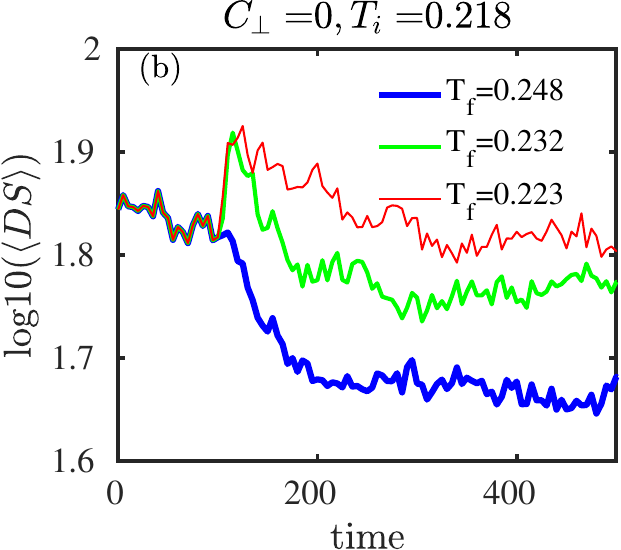}
\caption{\footnotesize {} Mean domain size $\langle DS \rangle$ of $q_{l,m}>0$ for (a) $C_{\perp}=0.05$ and (b) $C_{\perp}=0$. Dashed lines represent the fit $\sim t^{-\beta}$ and inset represents the values of $\beta$. The other parameters are $L=256$ and $M=10$. }
  \label{fig:domain_size}

    \includegraphics[scale=1.0,width=0.23\textwidth]{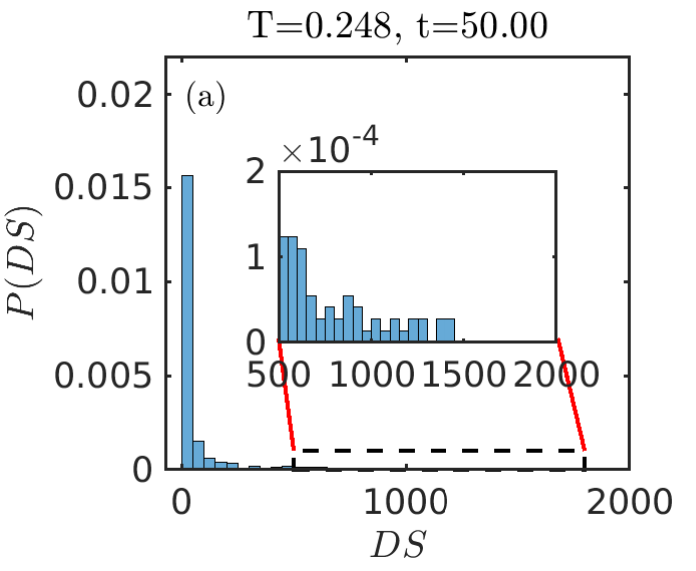}
       \includegraphics[scale=1.0,width=0.23\textwidth]{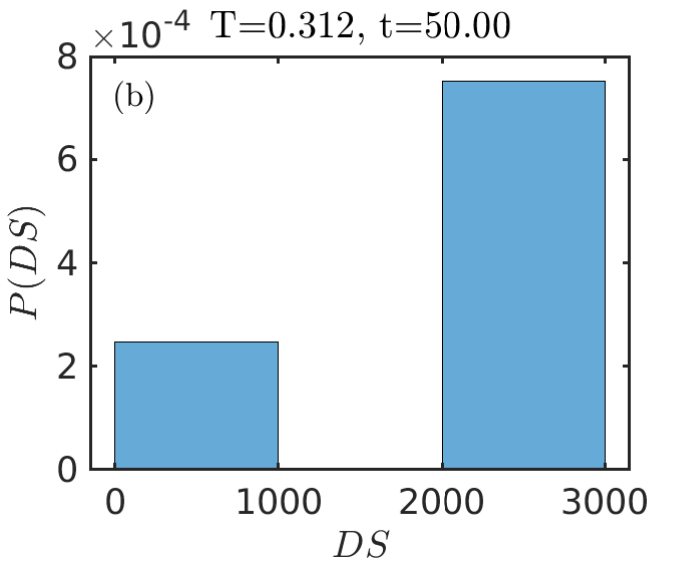}
\caption{\footnotesize {} Histogram  of domain size $DS$ of $q_{l,m}>0$ for (a) $C_{\perp}=0$ and (b) $C_{\perp}=0.05$ at a fixed time before the kick. The other parameters are $L=256$ and $M=10$. }
  \label{fig:hist_domain_size_initial}

    \includegraphics[scale=1.0,width=0.23\textwidth]{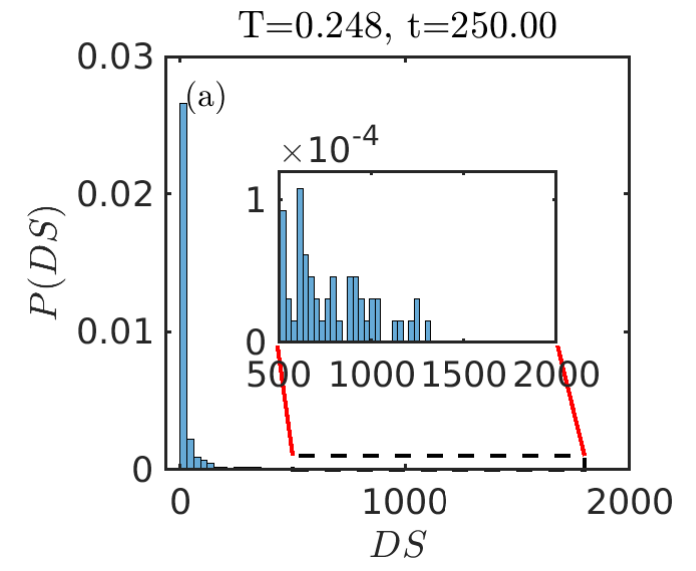}
       \includegraphics[scale=1.0,width=0.23\textwidth]{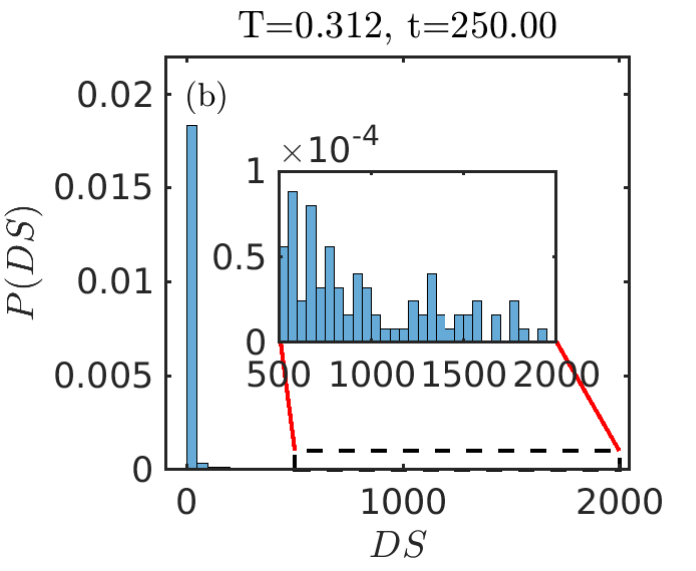}
\caption{\footnotesize {} Histogram  of domain size $DS$ of $q_{l,m}>0$ for (a) $C_{\perp}=0$ and (b) $C_{\perp}=0.05$ at a fixed time after the kick. The other parameters are $L=256$ and $M=10$. }
  \label{fig:hist_domain_size_final}
   \end{figure}

                    \begin{figure}[!htbp]   
    \includegraphics[scale=1.0,width=0.4\textwidth]{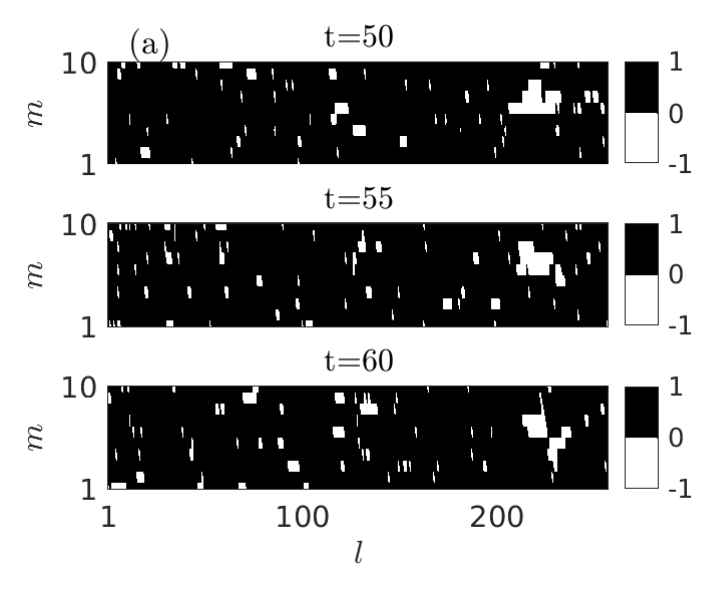}
     \includegraphics[scale=1.0,width=0.4\textwidth]{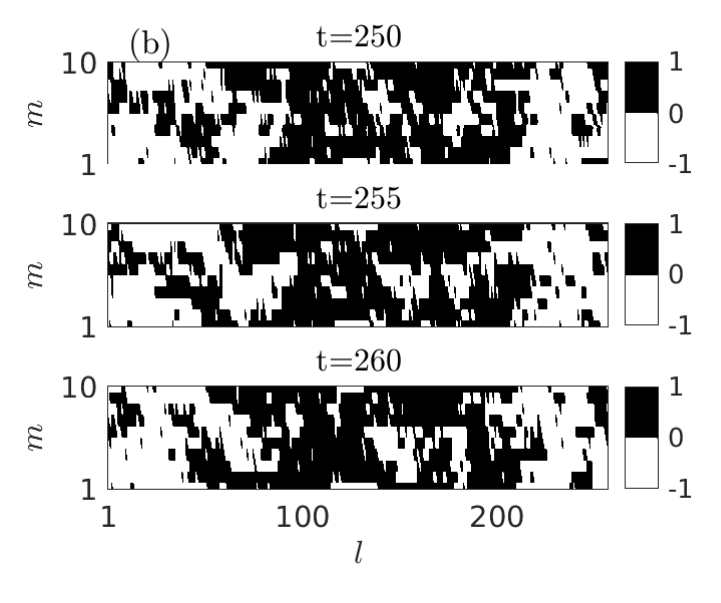}
       \includegraphics[scale=1.0,width=0.4\textwidth]{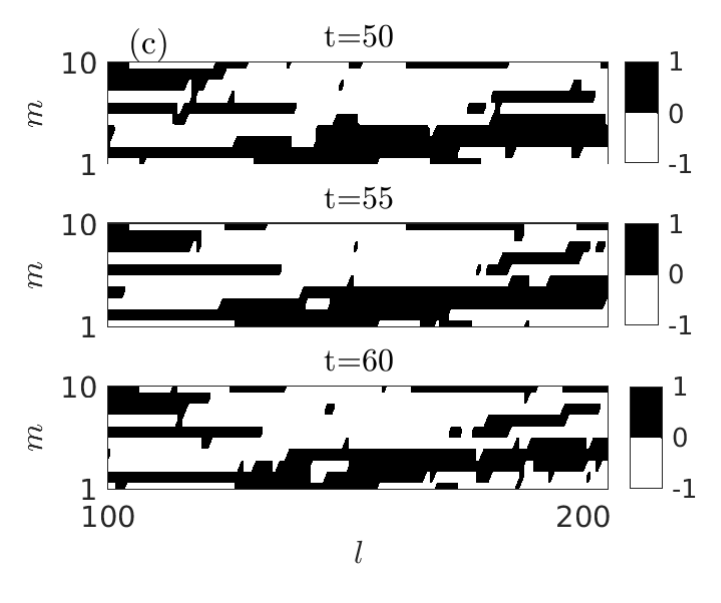}
\caption{\footnotesize {} Evolution of domain of $q_{l,m}>0$ for $C_{\perp}=0.05$ (triplets of (a) and (b)) and $C_{\perp}=0$  (triplets of (c))  before the kick. The other parameters are $L=256$ and $M=10$. }
  \label{fig:moving_kinks_multi}
   \end{figure}

         \begin{figure*}[!htbp]  
    \includegraphics[scale=1.0,width=0.32\textwidth]{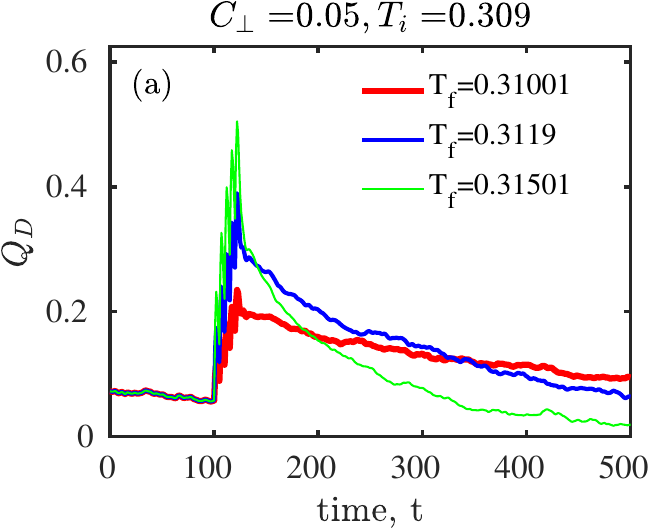}
    \includegraphics[scale=1.0,width=0.32\textwidth]{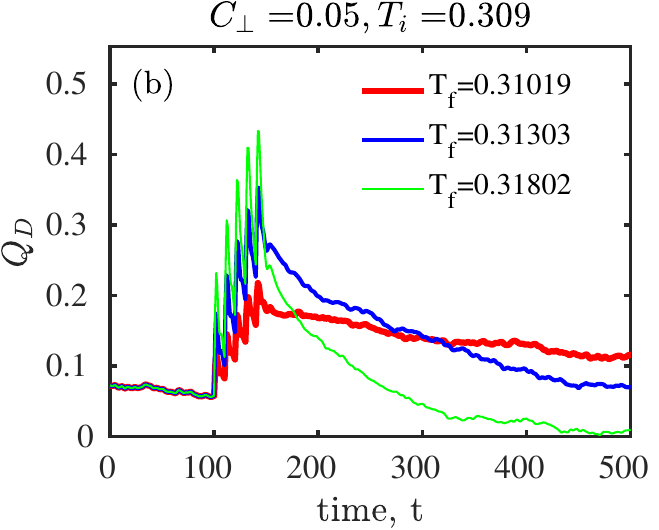}
    \includegraphics[scale=1.0,width=0.32\textwidth]{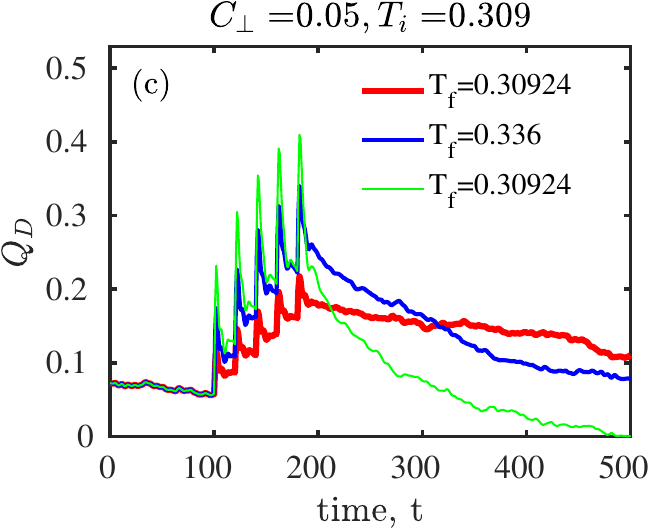}
\caption{\label{fig:ws}{\footnotesize {} Average displacement $Q_D$ of the multi-kick case for $C_{\perp}=0.05$, for three different kick intervals (a) 5, (b) 10, and (c) 20 time units, where $T_f$ represents the final temperature. The other parameters are $L=256$ and $M=10$. }}
  \label{fig:mkickd}
   \end{figure*}
   
         \begin{figure}[!htbp]  
\includegraphics[scale=1.0,width=0.4\textwidth]{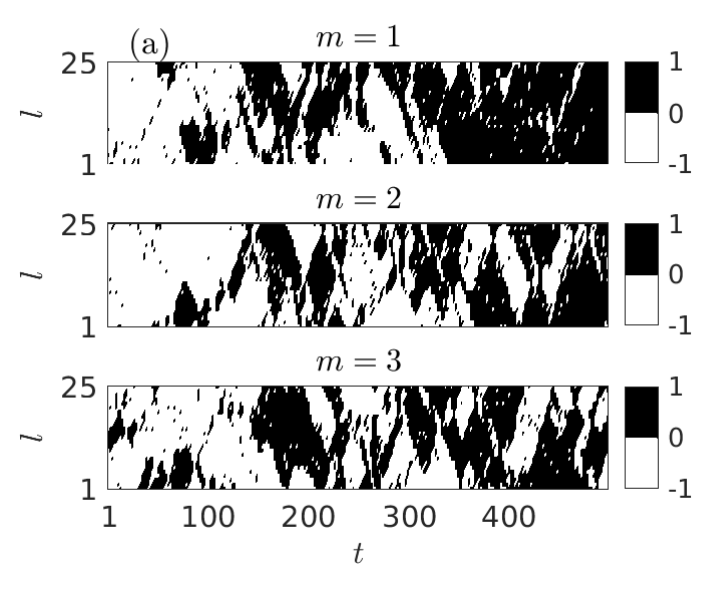}\\
 \includegraphics[scale=1.0,width=0.4\textwidth]{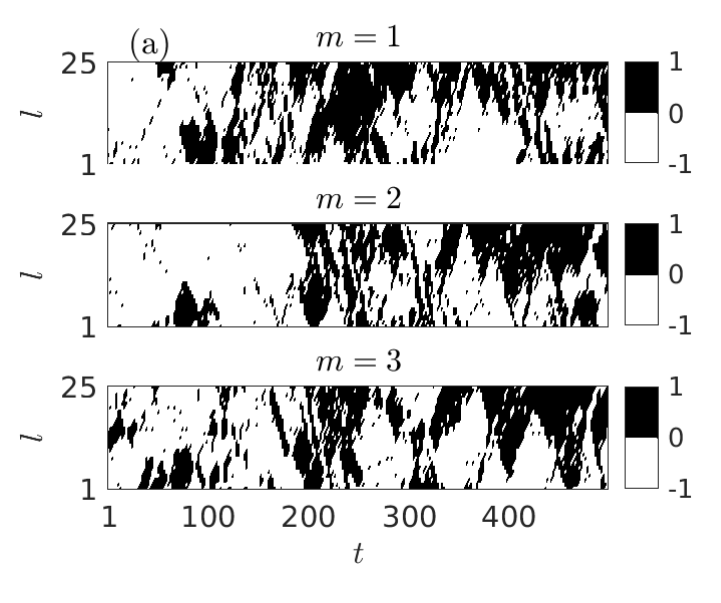}
\caption{{\footnotesize {} Soliton detector for multi-kick case: The triplets represent the cases of $C_{\perp}=0.05$ for two different kick strengths (a) $P_0=1$ and (b) $P_0=3$ for the fixed kick interval 20. Each of the triplets contains three chains. The other parameters are $C_{||}=1$, $L=256$ and $M=10$.  }}
  \label{fig:evolutionmulti}
   \end{figure}

            \begin{figure}[!htbp]  
    \includegraphics[scale=1.0,width=0.4\textwidth]{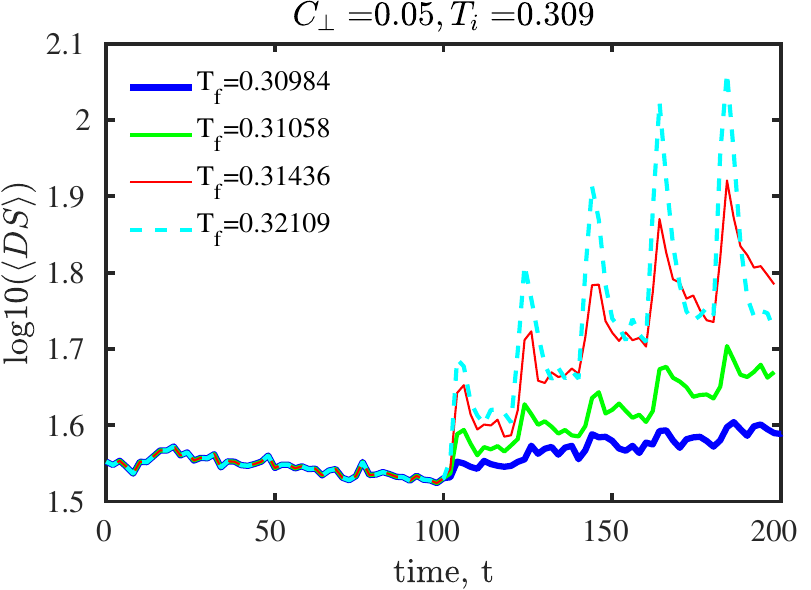}
\caption{\label{fig:ws}{\footnotesize {} Mean domain size of the multi-kick case for $C_{\perp}=0.05$, for four different kick strengths for a fixed kick interval 20, where $T_f$ represents the final temperature at the time $t=500$. The other parameters are $L=256$ and $M=10$.  }}
  \label{fig:mkickdomain}
   \end{figure}
 \section{Numerical Analysis of the model}
We study the dynamical evolution of the system for the initial condition $q^0_{l,m}=1$. We select $p^0_{l,m}$ from a uniform random distribution $[-0.5,0.5]$ and rescale it to preserve the conservation of $\mathcal{H}$. This rescaling is done by considering $p_{l,m}=\sqrt{\frac{\mathcal{H}-\mathcal{H}_V}{\mathcal{H}_K^0}}p^0_{l,m}$, where
  \begin{equation}
     \mathcal{H}_K^0=\sum_{l=1}^L \sum_{m=1}^M \frac{(p^0_{l,m})^2}{2}.
  \end{equation}
We then evolve the system for a pre-relaxation time $t=2000$ before starting to measure the observables considered below. 

We select $p^0_{l,m}$ from a uniform random distribution $[-0.5,0.5]$ and evolve the system for a pre-relaxation time $t=2000$ before starting to measure the observables considered below. Unless otherwise mentioned, we consider $L=256$ and $M=10$.

It is well known that for $C_{\perp}=0$, Eq.~\eqref{eq:phi4} approaches the continuum $\phi^4$
that supports both the kink and breather (kink-antikink bound state) solutions, where the energy of a kink is estimated to be 
 $$E_K=\frac{2\sqrt{2}}{3} \sqrt{\frac{C_{||}}{|A|}} \frac{|A|^2}{B}.$$ It has been found that the dynamical equations, Eq.~\eqref{eq:phi4}, also exhibit kink and transient breather excitations \cite{PhysRevB.34.6295}.
 
The structural transition associated with the model resulting from the underlying double well potential can be assessed with the observable time-averaged displacement $\langle Q_D \rangle$ defined as
\begin{equation}
\langle Q_D \rangle =\frac{1}{\tilde{T}}\int_0^{\tilde{T}}Q_D(t) dt,
\end{equation}
where  $Q_D= \frac{1}{L\times M}\sum_{l,m} q_{l,m}(t)$ and
$\tilde{T}$ denotes the time horizon of the numerical simulation
(the latter is not to be confused with the kinetic temperature
definition provided below). This structural transition is led by the (kinetic) temperature $T = \frac{1}{L\times M}\frac{1}{\tilde{T}} \int_0^{\tilde{T}} \sum_{l,m} v_{l,m}^2dt,$
where $v_{l,m}$ represents the velocity of the $(l,m)$th particle. 
Notice that here we are following the
definition of~\cite{PhysRevB.34.6295} (where the
customary factor of $2$ in the denominator is
not included).
Fig.~\ref{fig:pav} shows the time-averaged displacement as a function of temperature for two different inter-chain coupling strengths. The results are consistent with the analytical estimation that the critical temperature $T_c \rightarrow 0$ as $C_{\perp} \rightarrow 0$, and moreover the existence of a finite $T_c$ for  $C_{\perp} > 0$ \cite{PhysRevB.12.2824}; where $T_c$ is defined as the temperature
 at which the averaged displacement $\langle Q_D \rangle$ deviates from $0$.

We now examine the dynamical properties of the system by using spectral functions, while also reproducing the
results discussed in the earlier work of~\cite{PhysRevB.34.6295}. The spectral function
is defined as
\begin{equation}
    D(\mathbf{k},\omega)=\lim_{t_{max} \rightarrow \infty }\frac{1}{2 t_{max}} \frac{1}{L \times M}|q(\mathbf{k},\omega)|^2,
\end{equation}
where 
$q(\mathbf{k},\omega)=\int_{-t_{max}}^{t_{max}} dte^{-i \omega t}q(\mathbf{k},t)$ and $q(\mathbf{k},t)=\sum_{\mathbf{r}}e^{-i \mathbf{k}.\mathbf{r}}q(\mathbf{r},t)$. The spectral function is expected to show a peak corresponding to the frequency of particle motion across the double-well potential at higher temperatures \cite{BISHOP19801, PhysRevLett.42.937,PhysRevB.24.6566}.  On the other hand, at lower temperatures, this spectral function exhibits a peak corresponding to the frequency of the particle motion in a single well and an
additional peak resulting from K and $\bar{\rm K}$ diffusion. Fig.~\ref{fig:wsp}(a-b) shows the spectral function $D(k=0,\omega)$ for different values of $C_{\perp}$ at different temperatures $T$. We follow the peak that corresponds to the particle motion across the double well potential and observe the decrease in frequency as $T$ approaches $T_s$: this is called the soft-mode frequency $\omega_s$ (see Fig.~\ref{fig:wsp}(c)). As $T$ decreases below $T_s$, the frequency corresponds to oscillations in a single well and hardens again. Note the near-linear $T$ dependence of the squared frequency $\omega_s^2(k=0)$ as $T \rightarrow T_s$ from above and below, as predicted by mean-field (self-
consistent phonon) theory. We see that $T_c$, the critical phase transition temperature, is indeed greater than $T_s$, the dynamical crossover temperature

We now focus on the dynamical processes in the regime $T_c \lesssim T_s/3$, where the topological K, $\bar{\rm K}$ excitations dominate the  relaxation.  In what follows, we will either once or periodically perturb the system by imparting energy with instantaneous  kicks and will subsequently seek to measure relevant observables. A drive to the system with an amplitude $P_0$ with period $t_p$ leads to the modification of the equations of motion,
according to:
   \begin{align}
  \ddot{q}_{l,m}= & q_{l,m} - q_{l,m}^3+ C_{\perp} \big( q_{l+1, m}-2 q_{l,m}+ q_{l-1,m}) \nonumber\\ &  
 +C_{||}\big(q_{l, m+1}- 2 q_{l,m}+ q_{l, m-1} \Big)\nonumber \\
 &+P_0\delta(t-[t_0+nt_p]),~~n=0,1,2,...,
 \label{eq:phi4p}
\end{align}
 where $t_0$ is the initial offset time (not the pre-thermalization time). This offset time does not play any role in the system dynamics, and we fix $t_0=100$ for our analysis, unless otherwise mentioned, to see the difference in initial and final dynamics. 
\section{Single Kick Driving: $n=0$}
A single kick at $t_0$ changes the initial temperature $T_i$ of the system to the final temperature $T_f> T_c$. 
Recall that in both cases we refer to kinetic temperatures
in this statement.
The amplitude of kick strength $P_0$ determines the imparted
energy and accordingly affects the resulting kinetic temperature $T_f$. Fig.~\ref{fig:skickd} shows the evolution of the average displacement $Q_D(t)$ for different values of $C_{\perp}$. 
For a finite $C_{\perp}>0$, the slow equilibration time of $Q_D$ from a finite initial value to the final value of $Q_D$ 
decreases with the increase in kick strength, since stronger
kicks produce higher T and the equilibration is then controlled by spatially extended vibrations and large kink densities. Weaker kicks produce lower amplitude oscillations and associated kink densities. The relaxation process is evident in the measurement of time-averaged displacement $\langle Q_D \rangle$ shown in Fig.~\ref{fig:skickavgd}. We introduce a kink detector that detects the transition between wells  in the double well potential; see also the definition in \cite{PhysRevB.34.6295}. This diagnostic is used to
facilitate an understanding of the slow
relaxation. The kink detector results depicted in Fig.~\ref{fig:evolutionsingle} highlight the motion of the boundaries of black-white
regions, namely kink solitary waves. It is noticeable that, as compared with the plot of $C_{\perp}=0$ case of Fig.~\ref{fig:evolutioncperp0}, the kink detector at finite $C_{\perp}=0.05$ shows more travelling small black spots -- these are large amplitude breather-like excitations, {i.e.}, kink-antikink bound states \cite{PhysRevB.34.6295}. 
We notice that, following a kick, the average displacement $Q_D$ relaxes exponentially; {i.e.}, $Q_D \sim e^{-\alpha t}$, where $\alpha$ increases with increase in kick strength, see Fig.~\ref{fig:skickdfit}. We attribute this increase in $b$  to an increase in large amplitude excitations.

The observable average displacement quantifies the relaxation in the regime $T_c < T \lesssim T_s/3$. Nevertheless, this observable is less effective in characterizing the local dynamics of the system. Hence, we now consider another
observable, the mean domain size $\langle DS \rangle$ that quantifies the number of nearest neighbor particles in 
either left or right wells of the double well potential: {i.e.}, $q_{l,m}>0$ or $q_{l,m}<0$. We focus on the case $q_{l,m}>0$ since on average the mean domain size of $q_{l,m}>0$  is expected to be equal to that of $q_{l,m}<0$ for $T>T_c$. The maximum of the mean domain size $\langle DS \rangle_{\text{max}}=L \times M$. Fig.~\ref{fig:domain_size} shows the evolution of mean domain size of $q_{l,m}>0$ for {two different values of $C_{\perp}$}.

Unlike the case of the average displacement $Q_D$, the mean domain size exhibits an algebraic decay for a finite $C_{\perp}>0$. The histogram of mean domain size shows that for $C_{\perp}>0$, the probability of larger domain size is higher before a kick (Fig.~\ref{fig:hist_domain_size_initial}) as compared to the probability after the kick (Fig.~\ref{fig:hist_domain_size_final}). This is because the kick gives it a higher $T$, hence smaller domains, $Q_D$. On the other hand, for uncoupled chains, $C_{\perp}=0$, the probability of larger domain size is smaller
(in comparison to the coupled case) 
before and after kicking. For a finite inter-chain coupling, we see the kink movement across the chains, corresponding to the evolution of 2d domains through kink motion along their boundaries.
\section{Multi-Kick Driving: $n>0$}
A central aim of our study is to mimic a simplified information measurement protocol by pulsing the system periodically to examine memory and decoherence effects. To avoid additional complications of driving the system through the phase transition, in this case we start with $T_i >T_c$ but limit the amplitude and number, $n$, of kicks so that $T_f$ remains in the topological (kink) dominated regime $T\lesssim T_s/3$. Specifically, we restrict $T_f < T_s$ by fixing $n=4$ and considering a smaller strength of $P_0$. The final temperature $T_f$ is measured at the $(n+1)^{\text{th}}$ kick. Fig.~\ref{fig:mkickd} shows the average displacement $Q_D$ of a multi-kick case for three different values of kick periods. For a fixed $C_{\perp}$, we see a more rapid relaxation of $Q_D$ for a higher  kick strength. Moreover, in this case the energy equilibrates in the system before the relaxation for weaker kick strengths. This scenario is quite evident from the kink picture given in Fig.~\ref{fig:evolutionmulti}. 
Indeed, for higher kick strengths we see large numbers of both K, 
$\bar{\rm K}$ and kink-antikink bound states ({i.e.}, large amplitude breather excitations). {A further characterization, based on the mean domain size, exhibits a monotonic increase for all final temperatures $T_f$ beyond $T_0$ as shown in Fig.~\ref{fig:mkickdomain}}. 
 The characteristics of mean domain size further show the more rapid equilibration of the system for higher final temperatures. The overall picture is that, for large kick periods or weak kick strengths, the system equilibrates to a new $T$ before the next kick. However, for small periods or higher kick strengths the system cannot equilibrate between the kicks. One quantification of this is the difference between the temperature following a kick and just before the next kick and the equilibrated $T$ for that kick strength (see Fig.~\ref{fig:mkickd}). This suggests a crossover in periodicity-kick strength space from decoherence/loss of memory to finite coherence/some memory. It will be important to monitor this with information-theoretic measures (as in complex Ginzburg–Landau (CGL) equation case \cite{PhysRevE.105.034210}); since we expect the topological kink dynamics to primarily control this behavior, more detailed kink-following methods will be required. We believe that this is an
 important direction for further exploration.

   \section{Conclusions and Future Challenges}
   We have explored how weakly-coupled classical $\phi^4$ chains mimic relaxation and memory regimes exploited in quantum annealing architectures such as the one utilized
   for D-Wave based on TFI interactions through an analogy between the two classes of models. Indeed, this was motivated by an exact d-dimension to (d-1)-dimension model mapping between these two settings for equilibrium properties. Our considerations exemplify the importance for information processing of recognizing similarities between key phenomena exhibited by appropriate classical nonlinear equations and quantum linear ones. In particular, we emphasized the fundamental role of topological excitations (kinks and antikinks here). These are the analogues of Ising spin domain walls in the quantum TFI.  In general, the topic of statistics of topological defects, their relationship to information-theoretic measures such as entropy, and their response to perturbations are important frontiers \cite{thudiyangal2024universal}. Our findings pave the way for an understanding of the regimes where these analogies can be explored and the impact of the intensity and the time-duration of  the (measurement-emulating) perturbation towards the potential loss of information (effective decoherence) of the systems at hand.
        
   Our results can be extended in several directions. For example using different kick profiles (analogous to general Kibble-Zurek regimes as $T_c$ is crossed \cite{del2014universality}), introducing kink defects (analogous to bit errors), studying weakly-coupled chains in higher dimensions, other topologies, geometries, or inter-chain connectivities. The exact d to (d-1)-dimensional mapping leveraged above no longer exists in such cases, but we anticipate similar roles of topological excitations and their interplay with disorder. Finally, we have not included any dissipation mechanisms here, thus modeling a closed system. Including dissipation would extend the study to open systems, a topic of particular interest in its own right. The D-Wave quantum simulation of the 1d TFI model has been reported in \cite{king2023quantum} (see also~\cite{king2022}). From the perspective of the mapping to weakly-coupled classical $\phi^4$ chains discussed here, it would be interesting to experimentally study quasi-1d ferro-distortive materials \cite{collins1979dynamics}. For example subjecting them to periodic pulsing and monitoring displacements, dynamic structure factors and, even more importantly, 
   local (atomic displacements and electronic) properties could be monitored, e.g., with scanning tunneling probes or indirectly through nonlinear susceptibilities. Since the inter-chain coupling maps to the transverse field in the 1d TFI, that could be mimicked with a tranverse stress field. These are some among the many possible directions for nonlinear-to-quantum connections that are worthwhile considering in future work. Given the analogy of kink-kink coupling between two chains and electron diffusion in disordered systems (``trapping/detrapping") analytical progress might be possible by just considering two weakly coupled chains.

\acknowledgements This material is based upon work supported by the U.S.\ National Science 
Foundation under the awards   PHY-2110030 and DMS-2204702 (PGK).  
The work at LANL was carried out under the auspices of the U.S. DOE and 
NNSA
under Contract No. DEAC5206NA25396.

\bibliographystyle{apsrev4}
\let\itshape\upshape
\normalem
\bibliography{reference1}

\providecommand{\noopsort}[1]{}\providecommand{\singleletter}[1]{#1}%
\begin{thebibliography}{32}%
\makeatletter
\providecommand \@ifxundefined [1]{%
 \@ifx{#1\undefined}
}%
\providecommand \@ifnum [1]{%
 \ifnum #1\expandafter \@firstoftwo
 \else \expandafter \@secondoftwo
 \fi
}%
\providecommand \@ifx [1]{%
 \ifx #1\expandafter \@firstoftwo
 \else \expandafter \@secondoftwo
 \fi
}%
\providecommand \natexlab [1]{#1}%
\providecommand \enquote  [1]{``#1''}%
\providecommand \bibnamefont  [1]{#1}%
\providecommand \bibfnamefont [1]{#1}%
\providecommand \citenamefont [1]{#1}%
\providecommand \href@noop [0]{\@secondoftwo}%
\providecommand \href [0]{\begingroup \@sanitize@url \@href}%
\providecommand \@href[1]{\@@startlink{#1}\@@href}%
\providecommand \@@href[1]{\endgroup#1\@@endlink}%
\providecommand \@sanitize@url [0]{\catcode `\\12\catcode `\$12\catcode
  `\&12\catcode `\#12\catcode `\^12\catcode `\_12\catcode `\%12\relax}%
\providecommand \@@startlink[1]{}%
\providecommand \@@endlink[0]{}%
\providecommand \url  [0]{\begingroup\@sanitize@url \@url }%
\providecommand \@url [1]{\endgroup\@href {#1}{\urlprefix }}%
\providecommand \urlprefix  [0]{URL }%
\providecommand \Eprint [0]{\href }%
\providecommand \doibase [0]{http://dx.doi.org/}%
\providecommand \selectlanguage [0]{\@gobble}%
\providecommand \bibinfo  [0]{\@secondoftwo}%
\providecommand \bibfield  [0]{\@secondoftwo}%
\providecommand \translation [1]{[#1]}%
\providecommand \BibitemOpen [0]{}%
\providecommand \bibitemStop [0]{}%
\providecommand \bibitemNoStop [0]{.\EOS\space}%
\providecommand \EOS [0]{\spacefactor3000\relax}%
\providecommand \BibitemShut  [1]{\csname bibitem#1\endcsname}%
\let\auto@bib@innerbib\@empty
\bibitem [{\citenamefont {Kevrekidis}\ and\ \citenamefont
  {Cuevas-Maraver}(2019)}]{dkc}%
  \BibitemOpen
  \bibfield  {author} {\bibinfo {author} {\bibfnamefont {P.~G.}\ \bibnamefont
  {Kevrekidis}}\ and\ \bibinfo {author} {\bibfnamefont {J.}~\bibnamefont
  {Cuevas-Maraver}},\ }\href
  {https://link.springer.com/book/10.1007/978-3-030-11839-6} {\emph {\bibinfo
  {title} {{A Dynamical Perspective on the $\phi^4$ Model: Past, Present and
  Future}}}},\ \bibinfo {edition} {1st}\ ed.\ (\bibinfo  {publisher} {Springer
  Nature},\ \bibinfo {address} {Heidelberg},\ \bibinfo {year}
  {2019})\BibitemShut {NoStop}%
\bibitem [{\citenamefont {Venkataraman}(1979)}]{venkataraman1979soft}%
  \BibitemOpen
  \bibfield  {author} {\bibinfo {author} {\bibfnamefont {G.}~\bibnamefont
  {Venkataraman}},\ }\bibfield  {title} {\enquote {\bibinfo {title} {Soft modes
  and structural phase transitions},}\ }\href@noop {} {\bibfield  {journal}
  {\bibinfo  {journal} {Bulletin of Materials Science}\ }\textbf {\bibinfo
  {volume} {1}},\ \bibinfo {pages} {129} (\bibinfo {year} {1979})}\BibitemShut
  {NoStop}%
\bibitem [{\citenamefont {Howard}\ and\ \citenamefont
  {Stokes}(2005)}]{howard2005structures}%
  \BibitemOpen
  \bibfield  {author} {\bibinfo {author} {\bibfnamefont {C.~J.}\ \bibnamefont
  {Howard}}\ and\ \bibinfo {author} {\bibfnamefont {H.~T.}\ \bibnamefont
  {Stokes}},\ }\bibfield  {title} {\enquote {\bibinfo {title} {Structures and
  phase transitions in perovskites--a group-theoretical approach},}\
  }\href@noop {} {\bibfield  {journal} {\bibinfo  {journal} {Acta
  Crystallographica Section A: Foundations of Crystallography}\ }\textbf
  {\bibinfo {volume} {61}},\ \bibinfo {pages} {93} (\bibinfo {year}
  {2005})}\BibitemShut {NoStop}%
\bibitem [{\citenamefont {Whiticar}\ \emph {et~al.}(2023)\citenamefont
  {Whiticar}, \citenamefont {Smirnov}, \citenamefont {Lanting}, \citenamefont
  {Whittaker}, \citenamefont {Altomare}, \citenamefont {Medina}, \citenamefont
  {Deshpande}, \citenamefont {Ejtemaee}, \citenamefont {Hoskinson},
  \citenamefont {Babcock},\ and\ \citenamefont {Amin}}]{PhysRevB.107.075412}%
  \BibitemOpen
  \bibfield  {author} {\bibinfo {author} {\bibfnamefont {A.~M.}\ \bibnamefont
  {Whiticar}}, \bibinfo {author} {\bibfnamefont {A.~Y.}\ \bibnamefont
  {Smirnov}}, \bibinfo {author} {\bibfnamefont {T.}~\bibnamefont {Lanting}},
  \bibinfo {author} {\bibfnamefont {J.}~\bibnamefont {Whittaker}}, \bibinfo
  {author} {\bibfnamefont {F.}~\bibnamefont {Altomare}}, \bibinfo {author}
  {\bibfnamefont {T.}~\bibnamefont {Medina}}, \bibinfo {author} {\bibfnamefont
  {R.}~\bibnamefont {Deshpande}}, \bibinfo {author} {\bibfnamefont
  {S.}~\bibnamefont {Ejtemaee}}, \bibinfo {author} {\bibfnamefont
  {E.}~\bibnamefont {Hoskinson}}, \bibinfo {author} {\bibfnamefont
  {M.}~\bibnamefont {Babcock}}, \ and\ \bibinfo {author} {\bibfnamefont
  {M.~H.}\ \bibnamefont {Amin}},\ }\bibfield  {title} {\enquote {\bibinfo
  {title} {Probing flux and charge noise with macroscopic resonant
  tunneling},}\ }\href {\doibase 10.1103/PhysRevB.107.075412} {\bibfield
  {journal} {\bibinfo  {journal} {Phys. Rev. B}\ }\textbf {\bibinfo {volume}
  {107}},\ \bibinfo {pages} {075412} (\bibinfo {year} {2023})}\BibitemShut
  {NoStop}%
\bibitem [{\citenamefont {Dodd}\ \emph {et~al.}(1983)\citenamefont {Dodd},
  \citenamefont {Eilbeck}, \citenamefont {Gibbon},\ and\ \citenamefont
  {Morris}}]{Dodd}%
  \BibitemOpen
  \bibfield  {author} {\bibinfo {author} {\bibfnamefont {R.}~\bibnamefont
  {Dodd}}, \bibinfo {author} {\bibfnamefont {J.}~\bibnamefont {Eilbeck}},
  \bibinfo {author} {\bibfnamefont {J.}~\bibnamefont {Gibbon}}, \ and\ \bibinfo
  {author} {\bibfnamefont {H.}~\bibnamefont {Morris}},\ }\href@noop {} {\emph
  {\bibinfo {title} {Solitons and Nonlinear Wave Equations}}}\ (\bibinfo
  {publisher} {Academic},\ \bibinfo {address} {New York},\ \bibinfo {year}
  {1983})\BibitemShut {NoStop}%
\bibitem [{\citenamefont {Koehler}\ \emph {et~al.}(1975)\citenamefont
  {Koehler}, \citenamefont {Bishop}, \citenamefont {Krumhansl},\ and\
  \citenamefont {Schrieffer}}]{koehler1975molecular}%
  \BibitemOpen
  \bibfield  {author} {\bibinfo {author} {\bibfnamefont {T.}~\bibnamefont
  {Koehler}}, \bibinfo {author} {\bibfnamefont {A.}~\bibnamefont {Bishop}},
  \bibinfo {author} {\bibfnamefont {J.}~\bibnamefont {Krumhansl}}, \ and\
  \bibinfo {author} {\bibfnamefont {J.}~\bibnamefont {Schrieffer}},\ }\bibfield
   {title} {\enquote {\bibinfo {title} {Molecular dynamics simulation of a
  model for (one-dimensional) structural phase transitions},}\ }\href@noop {}
  {\bibfield  {journal} {\bibinfo  {journal} {Solid State Communications}\
  }\textbf {\bibinfo {volume} {17}},\ \bibinfo {pages} {1515} (\bibinfo {year}
  {1975})}\BibitemShut {NoStop}%
\bibitem [{\citenamefont {Kerr}\ and\ \citenamefont
  {Bishop}(1986)}]{PhysRevB.34.6295}%
  \BibitemOpen
  \bibfield  {author} {\bibinfo {author} {\bibfnamefont {W.~C.}\ \bibnamefont
  {Kerr}}\ and\ \bibinfo {author} {\bibfnamefont {A.~R.}\ \bibnamefont
  {Bishop}},\ }\bibfield  {title} {\enquote {\bibinfo {title} {Dynamics of
  structural phase transitions in highly anisotropic systems},}\ }\href
  {\doibase 10.1103/PhysRevB.34.6295} {\bibfield  {journal} {\bibinfo
  {journal} {Phys. Rev. B}\ }\textbf {\bibinfo {volume} {34}},\ \bibinfo
  {pages} {6295} (\bibinfo {year} {1986})}\BibitemShut {NoStop}%
\bibitem [{\citenamefont {Habib}\ and\ \citenamefont
  {Lythe}(2000)}]{PhysRevLett.84.1070}%
  \BibitemOpen
  \bibfield  {author} {\bibinfo {author} {\bibfnamefont {S.}~\bibnamefont
  {Habib}}\ and\ \bibinfo {author} {\bibfnamefont {G.}~\bibnamefont {Lythe}},\
  }\bibfield  {title} {\enquote {\bibinfo {title} {Dynamics of kinks:
  Nucleation, diffusion, and annihilation},}\ }\href {\doibase
  10.1103/PhysRevLett.84.1070} {\bibfield  {journal} {\bibinfo  {journal}
  {Phys. Rev. Lett.}\ }\textbf {\bibinfo {volume} {84}},\ \bibinfo {pages}
  {1070} (\bibinfo {year} {2000})}\BibitemShut {NoStop}%
\bibitem [{\citenamefont {Lythe}\ and\ \citenamefont {Habib}(2006)}]{1624301}%
  \BibitemOpen
  \bibfield  {author} {\bibinfo {author} {\bibfnamefont {G.}~\bibnamefont
  {Lythe}}\ and\ \bibinfo {author} {\bibfnamefont {S.}~\bibnamefont {Habib}},\
  }\bibfield  {title} {\enquote {\bibinfo {title} {Kink stochastics},}\ }\href
  {\doibase 10.1109/MCSE.2006.43} {\bibfield  {journal} {\bibinfo  {journal}
  {Computing in Science \& Engineering}\ }\textbf {\bibinfo {volume} {8}},\
  \bibinfo {pages} {10} (\bibinfo {year} {2006})}\BibitemShut {NoStop}%
\bibitem [{\citenamefont {Bishop}\ \emph {et~al.}(1989)\citenamefont {Bishop},
  \citenamefont {McLaughlin},\ and\ \citenamefont
  {Salerno}}]{bishop1989global}%
  \BibitemOpen
  \bibfield  {author} {\bibinfo {author} {\bibfnamefont {A.~R.}\ \bibnamefont
  {Bishop}}, \bibinfo {author} {\bibfnamefont {D.~W.}\ \bibnamefont
  {McLaughlin}}, \ and\ \bibinfo {author} {\bibfnamefont {M.}~\bibnamefont
  {Salerno}},\ }\bibfield  {title} {\enquote {\bibinfo {title} {Global
  coordinates for the breather-kink (antikink) sine-gordon phase space: An
  explicit separatrix as a possible source of chaos},}\ }\href {\doibase
  10.1103/PhysRevA.40.6463} {\bibfield  {journal} {\bibinfo  {journal} {Phys.
  Rev. A}\ }\textbf {\bibinfo {volume} {40}},\ \bibinfo {pages} {6463}
  (\bibinfo {year} {1989})}\BibitemShut {NoStop}%
\bibitem [{\citenamefont {Rasmussen}\ \emph {et~al.}(2000)\citenamefont
  {Rasmussen}, \citenamefont {Aubry}, \citenamefont {Bishop},\ and\
  \citenamefont {Tsironis}}]{rasmussen2000discrete}%
  \BibitemOpen
  \bibfield  {author} {\bibinfo {author} {\bibfnamefont {K.}~\bibnamefont
  {Rasmussen}}, \bibinfo {author} {\bibfnamefont {S.}~\bibnamefont {Aubry}},
  \bibinfo {author} {\bibfnamefont {A.}~\bibnamefont {Bishop}}, \ and\ \bibinfo
  {author} {\bibfnamefont {G.}~\bibnamefont {Tsironis}},\ }\bibfield  {title}
  {\enquote {\bibinfo {title} {Discrete nonlinear schr{\"o}dinger breathers in
  a phonon bath},}\ }\href@noop {} {\bibfield  {journal} {\bibinfo  {journal}
  {The European Physical Journal B-Condensed Matter and Complex Systems}\
  }\textbf {\bibinfo {volume} {15}},\ \bibinfo {pages} {169} (\bibinfo {year}
  {2000})}\BibitemShut {NoStop}%
\bibitem [{\citenamefont {Flach}\ and\ \citenamefont
  {Willis}(1998)}]{Flach_1998}%
  \BibitemOpen
  \bibfield  {author} {\bibinfo {author} {\bibfnamefont {S.}~\bibnamefont
  {Flach}}\ and\ \bibinfo {author} {\bibfnamefont {C.}~\bibnamefont {Willis}},\
  }\bibfield  {title} {\enquote {\bibinfo {title} {Discrete breathers},}\
  }\href {\doibase 10.1016/S0370-1573(97)00068-9} {\bibfield  {journal}
  {\bibinfo  {journal} {Phys. Rep.}\ }\textbf {\bibinfo {volume} {295}},\
  \bibinfo {pages} {181 } (\bibinfo {year} {1998})}\BibitemShut {NoStop}%
\bibitem [{\citenamefont {Flach}\ and\ \citenamefont
  {Gorbach}(2008)}]{Flach_2008}%
  \BibitemOpen
  \bibfield  {author} {\bibinfo {author} {\bibfnamefont {S.}~\bibnamefont
  {Flach}}\ and\ \bibinfo {author} {\bibfnamefont {A.~V.}\ \bibnamefont
  {Gorbach}},\ }\bibfield  {title} {\enquote {\bibinfo {title} {Discrete
  breathers - advances in theory and applications},}\ }\href {\doibase
  10.1016/j.physrep.2008.05.002} {\bibfield  {journal} {\bibinfo  {journal}
  {Phys. Rep.}\ }\textbf {\bibinfo {volume} {467}},\ \bibinfo {pages} {1 }
  (\bibinfo {year} {2008})}\BibitemShut {NoStop}%
\bibitem [{\citenamefont {Baker}\ \emph {et~al.}(1982)\citenamefont {Baker},
  \citenamefont {Bishop}, \citenamefont {Fesser}, \citenamefont {Beale},\ and\
  \citenamefont {Krumhansl}}]{Baker1982critical}%
  \BibitemOpen
  \bibfield  {author} {\bibinfo {author} {\bibfnamefont {G.~A.}\ \bibnamefont
  {Baker}}, \bibinfo {author} {\bibfnamefont {A.~R.}\ \bibnamefont {Bishop}},
  \bibinfo {author} {\bibfnamefont {K.}~\bibnamefont {Fesser}}, \bibinfo
  {author} {\bibfnamefont {P.~D.}\ \bibnamefont {Beale}}, \ and\ \bibinfo
  {author} {\bibfnamefont {J.~A.}\ \bibnamefont {Krumhansl}},\ }\bibfield
  {title} {\enquote {\bibinfo {title} {Critical and crossover behavior in the
  double-gaussian model on a lattice},}\ }\href {\doibase
  10.1103/PhysRevB.26.2596} {\bibfield  {journal} {\bibinfo  {journal} {Phys.
  Rev. B}\ }\textbf {\bibinfo {volume} {26}},\ \bibinfo {pages} {2596}
  (\bibinfo {year} {1982})}\BibitemShut {NoStop}%
\bibitem [{\citenamefont {Pfeuty}(1970)}]{pfeuty1970one}%
  \BibitemOpen
  \bibfield  {author} {\bibinfo {author} {\bibfnamefont {P.}~\bibnamefont
  {Pfeuty}},\ }\bibfield  {title} {\enquote {\bibinfo {title} {The
  one-dimensional ising model with a transverse field},}\ }\href@noop {}
  {\bibfield  {journal} {\bibinfo  {journal} {ANNALS of Physics}\ }\textbf
  {\bibinfo {volume} {57}},\ \bibinfo {pages} {79} (\bibinfo {year}
  {1970})}\BibitemShut {NoStop}%
\bibitem [{\citenamefont {Bernal~Neira}(2022)}]{bernal2022coherent}%
  \BibitemOpen
  \bibfield  {author} {\bibinfo {author} {\bibfnamefont {D.}~\bibnamefont
  {Bernal~Neira}},\ }\bibfield  {title} {\enquote {\bibinfo {title} {Coherent
  simulation with thousands of qubits},}\ }\href@noop {} {\bibfield  {journal}
  {\bibinfo  {journal} {Nature Physics}\ }\textbf {\bibinfo {volume} {18}},\
  \bibinfo {pages} {1273} (\bibinfo {year} {2022})}\BibitemShut {NoStop}%
\bibitem [{\citenamefont {King}\ \emph {et~al.}(2022)\citenamefont {King},
  \citenamefont {Suzuki}, \citenamefont {Raymond}, \citenamefont {Zucca},
  \citenamefont {Lanting}, \citenamefont {Altomare}, \citenamefont {Berkley},
  \citenamefont {Ejtemaee}, \citenamefont {Hoskinson}, \citenamefont {Huang},
  \citenamefont {Ladizinsky}, \citenamefont {MacDonald}, \citenamefont
  {Marsden}, \citenamefont {Oh}, \citenamefont {Poulin-Lamarre}, \citenamefont
  {Reis}, \citenamefont {Rich}, \citenamefont {Sato}, \citenamefont
  {Whittaker}, \citenamefont {Yao}, \citenamefont {Harris}, \citenamefont
  {Lidar}, \citenamefont {Nishimori},\ and\ \citenamefont {Amin}}]{king2022}%
  \BibitemOpen
  \bibfield  {author} {\bibinfo {author} {\bibfnamefont {A.~D.}\ \bibnamefont
  {King}}, \bibinfo {author} {\bibfnamefont {S.}~\bibnamefont {Suzuki}},
  \bibinfo {author} {\bibfnamefont {J.}~\bibnamefont {Raymond}}, \bibinfo
  {author} {\bibfnamefont {A.}~\bibnamefont {Zucca}}, \bibinfo {author}
  {\bibfnamefont {T.}~\bibnamefont {Lanting}}, \bibinfo {author} {\bibfnamefont
  {F.}~\bibnamefont {Altomare}}, \bibinfo {author} {\bibfnamefont {A.~J.}\
  \bibnamefont {Berkley}}, \bibinfo {author} {\bibfnamefont {S.}~\bibnamefont
  {Ejtemaee}}, \bibinfo {author} {\bibfnamefont {E.}~\bibnamefont {Hoskinson}},
  \bibinfo {author} {\bibfnamefont {S.}~\bibnamefont {Huang}}, \bibinfo
  {author} {\bibfnamefont {E.}~\bibnamefont {Ladizinsky}}, \bibinfo {author}
  {\bibfnamefont {A.~J.~R.}\ \bibnamefont {MacDonald}}, \bibinfo {author}
  {\bibfnamefont {G.}~\bibnamefont {Marsden}}, \bibinfo {author} {\bibfnamefont
  {T.}~\bibnamefont {Oh}}, \bibinfo {author} {\bibfnamefont {G.}~\bibnamefont
  {Poulin-Lamarre}}, \bibinfo {author} {\bibfnamefont {M.}~\bibnamefont
  {Reis}}, \bibinfo {author} {\bibfnamefont {C.}~\bibnamefont {Rich}}, \bibinfo
  {author} {\bibfnamefont {Y.}~\bibnamefont {Sato}}, \bibinfo {author}
  {\bibfnamefont {J.~D.}\ \bibnamefont {Whittaker}}, \bibinfo {author}
  {\bibfnamefont {J.}~\bibnamefont {Yao}}, \bibinfo {author} {\bibfnamefont
  {R.}~\bibnamefont {Harris}}, \bibinfo {author} {\bibfnamefont {D.~A.}\
  \bibnamefont {Lidar}}, \bibinfo {author} {\bibfnamefont {H.}~\bibnamefont
  {Nishimori}}, \ and\ \bibinfo {author} {\bibfnamefont {M.~H.}\ \bibnamefont
  {Amin}},\ }\bibfield  {title} {\enquote {\bibinfo {title} {Coherent quantum
  annealing in a programmable 2,000 qubit ising chain},}\ }\href {\doibase
  10.1038/s41567-022-01741-6} {\bibfield  {journal} {\bibinfo  {journal}
  {Nature Physics}\ }\textbf {\bibinfo {volume} {18}},\ \bibinfo {pages} {1324}
  (\bibinfo {year} {2022})}\BibitemShut {NoStop}%
\bibitem [{\citenamefont {Mithun}\ \emph {et~al.}(2022)\citenamefont {Mithun},
  \citenamefont {Kevrekidis}, \citenamefont {Saxena},\ and\ \citenamefont
  {Bishop}}]{PhysRevE.105.034210}%
  \BibitemOpen
  \bibfield  {author} {\bibinfo {author} {\bibfnamefont {T.}~\bibnamefont
  {Mithun}}, \bibinfo {author} {\bibfnamefont {P.~G.}\ \bibnamefont
  {Kevrekidis}}, \bibinfo {author} {\bibfnamefont {A.}~\bibnamefont {Saxena}},
  \ and\ \bibinfo {author} {\bibfnamefont {A.~R.}\ \bibnamefont {Bishop}},\
  }\bibfield  {title} {\enquote {\bibinfo {title} {Measurement and memory in
  the periodically driven complex ginzburg-landau equation},}\ }\href {\doibase
  10.1103/PhysRevE.105.034210} {\bibfield  {journal} {\bibinfo  {journal}
  {Phys. Rev. E}\ }\textbf {\bibinfo {volume} {105}},\ \bibinfo {pages}
  {034210} (\bibinfo {year} {2022})}\BibitemShut {NoStop}%
\bibitem [{\citenamefont {Lloyd}\ and\ \citenamefont
  {Slotine}(2000)}]{Lloyd2000quantum}%
  \BibitemOpen
  \bibfield  {author} {\bibinfo {author} {\bibfnamefont {S.}~\bibnamefont
  {Lloyd}}\ and\ \bibinfo {author} {\bibfnamefont {J.-J.~E.}\ \bibnamefont
  {Slotine}},\ }\bibfield  {title} {\enquote {\bibinfo {title} {Quantum
  feedback with weak measurements},}\ }\href {\doibase
  10.1103/PhysRevA.62.012307} {\bibfield  {journal} {\bibinfo  {journal} {Phys.
  Rev. A}\ }\textbf {\bibinfo {volume} {62}},\ \bibinfo {pages} {012307}
  (\bibinfo {year} {2000})}\BibitemShut {NoStop}%
\bibitem [{\citenamefont {Gr{\o}nbech-Jensen}\ \emph
  {et~al.}(1992)\citenamefont {Gr{\o}nbech-Jensen}, \citenamefont {Bishop},
  \citenamefont {Falo},\ and\ \citenamefont {Lomdahl}}]{gro1992langevin}%
  \BibitemOpen
  \bibfield  {author} {\bibinfo {author} {\bibfnamefont {N.}~\bibnamefont
  {Gr{\o}nbech-Jensen}}, \bibinfo {author} {\bibfnamefont {A.~R.}\ \bibnamefont
  {Bishop}}, \bibinfo {author} {\bibfnamefont {F.}~\bibnamefont {Falo}}, \ and\
  \bibinfo {author} {\bibfnamefont {P.~S.}\ \bibnamefont {Lomdahl}},\
  }\bibfield  {title} {\enquote {\bibinfo {title} {Langevin-dynamics simulation
  of relaxation in large frustrated josephson-junction arrays},}\ }\href
  {\doibase 10.1103/PhysRevB.45.10139} {\bibfield  {journal} {\bibinfo
  {journal} {Phys. Rev. B}\ }\textbf {\bibinfo {volume} {45}},\ \bibinfo
  {pages} {10139} (\bibinfo {year} {1992})}\BibitemShut {NoStop}%
\bibitem [{\citenamefont {Horovitz}\ \emph {et~al.}(1977)\citenamefont
  {Horovitz}, \citenamefont {Krumhansl},\ and\ \citenamefont
  {Domany}}]{horovitz1977solitons}%
  \BibitemOpen
  \bibfield  {author} {\bibinfo {author} {\bibfnamefont {B.}~\bibnamefont
  {Horovitz}}, \bibinfo {author} {\bibfnamefont {J.}~\bibnamefont {Krumhansl}},
  \ and\ \bibinfo {author} {\bibfnamefont {E.}~\bibnamefont {Domany}},\
  }\bibfield  {title} {\enquote {\bibinfo {title} {Solitons in a coupled linear
  chain system},}\ }\href@noop {} {\bibfield  {journal} {\bibinfo  {journal}
  {Physical Review Letters}\ }\textbf {\bibinfo {volume} {38}},\ \bibinfo
  {pages} {778} (\bibinfo {year} {1977})}\BibitemShut {NoStop}%
\bibitem [{\citenamefont {Aranson}\ and\ \citenamefont
  {Kramer}(2002)}]{aranson}%
  \BibitemOpen
  \bibfield  {author} {\bibinfo {author} {\bibfnamefont {I.~S.}\ \bibnamefont
  {Aranson}}\ and\ \bibinfo {author} {\bibfnamefont {L.}~\bibnamefont
  {Kramer}},\ }\bibfield  {title} {\enquote {\bibinfo {title} {The world of the
  complex ginzburg-landau equation},}\ }\href {\doibase
  10.1103/RevModPhys.74.99} {\bibfield  {journal} {\bibinfo  {journal} {Rev.
  Mod. Phys.}\ }\textbf {\bibinfo {volume} {74}},\ \bibinfo {pages} {99}
  (\bibinfo {year} {2002})}\BibitemShut {NoStop}%
\bibitem [{\citenamefont {LeSar}(2014)}]{LeSar2014simulations}%
  \BibitemOpen
  \bibfield  {author} {\bibinfo {author} {\bibfnamefont {R.}~\bibnamefont
  {LeSar}},\ }\bibfield  {title} {\enquote {\bibinfo {title} {Simulations of
  dislocation structure and response},}\ }\href {\doibase
  10.1146/annurev-conmatphys-031113-133858} {\bibfield  {journal} {\bibinfo
  {journal} {Annual Review of Condensed Matter Physics}\ }\textbf {\bibinfo
  {volume} {5}},\ \bibinfo {pages} {375} (\bibinfo {year} {2014})}\BibitemShut
  {NoStop}%
\bibitem [{\citenamefont {Nisoli}(2020)}]{Nisoli2020equilibrium}%
  \BibitemOpen
  \bibfield  {author} {\bibinfo {author} {\bibfnamefont {C.}~\bibnamefont
  {Nisoli}},\ }\bibfield  {title} {\enquote {\bibinfo {title} {Equilibrium
  field theory of magnetic monopoles in degenerate square spin ice:
  Correlations, entropic interactions, and charge screening regimes},}\ }\href
  {\doibase 10.1103/PhysRevB.102.220401} {\bibfield  {journal} {\bibinfo
  {journal} {Phys. Rev. B}\ }\textbf {\bibinfo {volume} {102}},\ \bibinfo
  {pages} {220401} (\bibinfo {year} {2020})}\BibitemShut {NoStop}%
\bibitem [{\citenamefont {Stoll}\ \emph {et~al.}(1979)\citenamefont {Stoll},
  \citenamefont {Schneider},\ and\ \citenamefont
  {Bishop}}]{PhysRevLett.42.937}%
  \BibitemOpen
  \bibfield  {author} {\bibinfo {author} {\bibfnamefont {E.}~\bibnamefont
  {Stoll}}, \bibinfo {author} {\bibfnamefont {T.}~\bibnamefont {Schneider}}, \
  and\ \bibinfo {author} {\bibfnamefont {A.~R.}\ \bibnamefont {Bishop}},\
  }\bibfield  {title} {\enquote {\bibinfo {title} {Evidence for breather
  excitations in the sine-gordon chain},}\ }\href {\doibase
  10.1103/PhysRevLett.42.937} {\bibfield  {journal} {\bibinfo  {journal} {Phys.
  Rev. Lett.}\ }\textbf {\bibinfo {volume} {42}},\ \bibinfo {pages} {937}
  (\bibinfo {year} {1979})}\BibitemShut {NoStop}%
\bibitem [{\citenamefont {Bishop}\ and\ \citenamefont
  {Krumhansl}(1975)}]{PhysRevB.12.2824}%
  \BibitemOpen
  \bibfield  {author} {\bibinfo {author} {\bibfnamefont {A.~R.}\ \bibnamefont
  {Bishop}}\ and\ \bibinfo {author} {\bibfnamefont {J.~A.}\ \bibnamefont
  {Krumhansl}},\ }\bibfield  {title} {\enquote {\bibinfo {title} {Mean field
  and exact results for structural phase transitions in one-dimensional and
  very anisotropic two-dimensional and three-dimensional systems},}\ }\href
  {\doibase 10.1103/PhysRevB.12.2824} {\bibfield  {journal} {\bibinfo
  {journal} {Phys. Rev. B}\ }\textbf {\bibinfo {volume} {12}},\ \bibinfo
  {pages} {2824} (\bibinfo {year} {1975})}\BibitemShut {NoStop}%
\bibitem [{\citenamefont {Bishop}\ \emph {et~al.}(1980)\citenamefont {Bishop},
  \citenamefont {Krumhansl},\ and\ \citenamefont {Trullinger}}]{BISHOP19801}%
  \BibitemOpen
  \bibfield  {author} {\bibinfo {author} {\bibfnamefont {A.}~\bibnamefont
  {Bishop}}, \bibinfo {author} {\bibfnamefont {J.}~\bibnamefont {Krumhansl}}, \
  and\ \bibinfo {author} {\bibfnamefont {S.}~\bibnamefont {Trullinger}},\
  }\bibfield  {title} {\enquote {\bibinfo {title} {Solitons in condensed
  matter: A paradigm},}\ }\href {\doibase
  https://doi.org/10.1016/0167-2789(80)90003-2} {\bibfield  {journal} {\bibinfo
   {journal} {Physica D: Nonlinear Phenomena}\ }\textbf {\bibinfo {volume}
  {1}},\ \bibinfo {pages} {1} (\bibinfo {year} {1980})}\BibitemShut {NoStop}%
\bibitem [{\citenamefont {Kerr}\ \emph {et~al.}(1981)\citenamefont {Kerr},
  \citenamefont {Baeriswyl},\ and\ \citenamefont {Bishop}}]{PhysRevB.24.6566}%
  \BibitemOpen
  \bibfield  {author} {\bibinfo {author} {\bibfnamefont {W.~C.}\ \bibnamefont
  {Kerr}}, \bibinfo {author} {\bibfnamefont {D.}~\bibnamefont {Baeriswyl}}, \
  and\ \bibinfo {author} {\bibfnamefont {A.~R.}\ \bibnamefont {Bishop}},\
  }\bibfield  {title} {\enquote {\bibinfo {title} {Equilibrium dynamics of the
  sine-gordon chain: A molecular-dynamics study},}\ }\href {\doibase
  10.1103/PhysRevB.24.6566} {\bibfield  {journal} {\bibinfo  {journal} {Phys.
  Rev. B}\ }\textbf {\bibinfo {volume} {24}},\ \bibinfo {pages} {6566}
  (\bibinfo {year} {1981})}\BibitemShut {NoStop}%
\bibitem [{\citenamefont {Thudiyangal}\ and\ \citenamefont {del
  Campo}(2024)}]{thudiyangal2024universal}%
  \BibitemOpen
  \bibfield  {author} {\bibinfo {author} {\bibfnamefont {M.}~\bibnamefont
  {Thudiyangal}}\ and\ \bibinfo {author} {\bibfnamefont {A.}~\bibnamefont {del
  Campo}},\ }\href@noop {} {\enquote {\bibinfo {title} {Universal vortex
  statistics and stochastic geometry of bose-einstein condensation},}\ }
  (\bibinfo {year} {2024}),\ \Eprint {http://arxiv.org/abs/2401.09525}
  {arXiv:2401.09525 [cond-mat.quant-gas]} \BibitemShut {NoStop}%
\bibitem [{\citenamefont {Del~Campo}\ and\ \citenamefont
  {Zurek}(2014)}]{del2014universality}%
  \BibitemOpen
  \bibfield  {author} {\bibinfo {author} {\bibfnamefont {A.}~\bibnamefont
  {Del~Campo}}\ and\ \bibinfo {author} {\bibfnamefont {W.~H.}\ \bibnamefont
  {Zurek}},\ }\bibfield  {title} {\enquote {\bibinfo {title} {Universality of
  phase transition dynamics: Topological defects from symmetry breaking},}\
  }\href@noop {} {\bibfield  {journal} {\bibinfo  {journal} {International
  Journal of Modern Physics A}\ }\textbf {\bibinfo {volume} {29}},\ \bibinfo
  {pages} {1430018} (\bibinfo {year} {2014})}\BibitemShut {NoStop}%
\bibitem [{\citenamefont {King}\ \emph {et~al.}(2023)\citenamefont {King},
  \citenamefont {Raymond}, \citenamefont {Lanting}, \citenamefont {Harris},
  \citenamefont {Zucca}, \citenamefont {Altomare}, \citenamefont {Berkley},
  \citenamefont {Boothby}, \citenamefont {Ejtemaee}, \citenamefont {Enderud}
  \emph {et~al.}}]{king2023quantum}%
  \BibitemOpen
  \bibfield  {author} {\bibinfo {author} {\bibfnamefont {A.~D.}\ \bibnamefont
  {King}}, \bibinfo {author} {\bibfnamefont {J.}~\bibnamefont {Raymond}},
  \bibinfo {author} {\bibfnamefont {T.}~\bibnamefont {Lanting}}, \bibinfo
  {author} {\bibfnamefont {R.}~\bibnamefont {Harris}}, \bibinfo {author}
  {\bibfnamefont {A.}~\bibnamefont {Zucca}}, \bibinfo {author} {\bibfnamefont
  {F.}~\bibnamefont {Altomare}}, \bibinfo {author} {\bibfnamefont {A.~J.}\
  \bibnamefont {Berkley}}, \bibinfo {author} {\bibfnamefont {K.}~\bibnamefont
  {Boothby}}, \bibinfo {author} {\bibfnamefont {S.}~\bibnamefont {Ejtemaee}},
  \bibinfo {author} {\bibfnamefont {C.}~\bibnamefont {Enderud}},  \emph
  {et~al.},\ }\bibfield  {title} {\enquote {\bibinfo {title} {Quantum critical
  dynamics in a 5,000-qubit programmable spin glass},}\ }\href@noop {}
  {\bibfield  {journal} {\bibinfo  {journal} {Nature}\ }\textbf {\bibinfo
  {volume} {617}},\ \bibinfo {pages} {61–66} (\bibinfo {year}
  {2023})}\BibitemShut {NoStop}%
\bibitem [{\citenamefont {Collins}\ \emph {et~al.}(1979)\citenamefont
  {Collins}, \citenamefont {Blumen}, \citenamefont {Currie},\ and\
  \citenamefont {Ross}}]{collins1979dynamics}%
  \BibitemOpen
  \bibfield  {author} {\bibinfo {author} {\bibfnamefont {M.}~\bibnamefont
  {Collins}}, \bibinfo {author} {\bibfnamefont {A.}~\bibnamefont {Blumen}},
  \bibinfo {author} {\bibfnamefont {J.}~\bibnamefont {Currie}}, \ and\ \bibinfo
  {author} {\bibfnamefont {J.}~\bibnamefont {Ross}},\ }\bibfield  {title}
  {\enquote {\bibinfo {title} {Dynamics of domain walls in ferrodistortive
  materials. i. theory},}\ }\href@noop {} {\bibfield  {journal} {\bibinfo
  {journal} {Physical Review B}\ }\textbf {\bibinfo {volume} {19}},\ \bibinfo
  {pages} {3630} (\bibinfo {year} {1979})}\BibitemShut {NoStop}%
\end{thebibliography}%




\end{document}